\documentclass{llncs}
\usepackage{amsfonts}
\usepackage{amssymb}
\usepackage{latexsym}
\usepackage{graphicx}

\begin{document}
	
	\title{A Foreground-Background queueing model with speed or 
		capacity modulation} 
	\author{
		Andrea Marin\inst{1} \and Isi Mitrani\inst{2}}

	\institute{
		Universit\`a Ca' Foscari Venezia, DAIS, Venice, Italy,
		\email{marin@unive.it}
		\and
		Newcastle University, School of Computing, UK\\
		\email{isi.mitrani@newcastle.ac.uk}
	}

\maketitle

\begin{abstract}
The models studied in the steady state involve two queues which are served 
either by a single server whose speed depends on the number of jobs present, 
or by several parallel servers whose number may be controlled dynamically. 
Job service times have a two-phase Coxian distribution and the second phase 
is given lower priority than the first. The trade-offs between holding costs and 
energy consumption costs are examined by means of a suitable cost functions. 
Two different two-dimensional Markov process are solved exactly. The solutions 
are used in several numerical experiments. Some counter-intuitive results are 
observed.
\end{abstract}

\section{Introduction}

The topic of controlling the energy consumption of computers has recived 
considerable attention in recent years. Some modern processors are designed 
so that the frequency at which they work, and hence the speed at which they 
execute jobs, can be adjusted dynamically depending on the number of jobs 
present. Several discrete frequency levels are supported, covering a wide range 
of possible speeds. The idea is that at light loads the processor would work at 
lower speed, reducing the energy costs, while at high loads it would speed up, 
reducing the job holding costs. This approach will be referred to as `speed 
modulation'. 

A similar energy-saving technique can be applied in traditional multiprocessor 
systems. Rather than modulating the speed of individual processors, one could 
control the number of processors that are working: switch one or more of them 
off during periods of light loads and back on when the load increases. This will 
be referred to as `capacity modulation'.

We are interested in evaluating the trade-offs arising in connection with both 
speed modulaton and capacity modulation, with a view to computing optimal 
operating policies.

Another factor that influences system performance is the job scheduling strategy. 
The commonly used First-Come-First-Served policy (FCFS, also called First-In-First-Out, 
or FIFO) performs well when the service times are not very variable, but it is far from 
optimal when the coefficient of variation is  high. In the latter case, it is well known 
that policies which favour short jobs over long ones have lower holding costs. In the 
case of a single processor, it was demonstrated by Schrage \cite{sch2} that the 
globally optimal scheduling strategy is Shortest-Remaining-Processing-Time-first 
(SRPT). However, that is not a practical policy because the exact processing times 
of incoming jobs are not usually known in advance. Two `blind' policies (i.e., they do 
not require advance knowledge of processing times) which favour short jobs are Processor-Sharing (PS) and Least-Attained-Service (LAS, also known as 
Foreground-Background, FB). Indeed, it was shown by Yashkov \cite{yas} that, 
among all blind policies and processing times distributions with increasing failure 
rate, LAS minimizes the average number of jobs in the system. 

Unfortunately, neither PS nor LAS are implementable in their pure form because they 
require more than one job to be served in parallel, which implies excessive levels of 
context switching. Moreover, LAS requires a dynamic number of priority queues to 
store the jobs that have received the same amount of service. 

Our aim is to study speed and capacity modulation in the context of a particular 
scheduling strategy that gives priority to shorter jobs without needing to keep track 
of their elapsed times. Two queues are employed, referred to as the foreground 
queue and the background queue. Jobs consist of one or two sequential phases, the 
first of which is executed in the foreground queue and the second, at lower priority, 
in the background queue. In this setting, instead of keeping track of the attained 
service of each job, we use the passage from the first to the second phase as the 
event that triggers the downgrading of the job priority. Clearly, the practical 
assumption is that the system is able to detect the change of the service phase. In 
fact, jobs consisting of more than one task are common in many areas of computer 
science. 

An example of an application that would lend itself to such a scheduling policy is 
one where commercial transactions access a database. All transacrions start by 
executing a task involving `read' queries. In many cases that is all they do before 
terminating. Alternatively, the read phase may be followed by an `update' phase which 
is more complex and takes longer to execute. Our scheduling strategy would place the 
read tasks in the foreground queue and the update ones in the background queue. 
Moreover, depending on whether we are dealing with a single speed-modulated 
processor, or a capacity-modulated multiprocessor system the service offered
would depend on the total number of jobs present.

To the best of our knowledge, such models have not been analysed before. Under 
appropriate assumptions, we obtain exact solutions for the two-dimensional 
Markov processes that describe the steady-state behaviour of the two queues. 
The single and multiprocessor models present distinct challenges in the analysis. 
In the case of a single processor, the background queue can be served only when 
the foreground queue is empty, whereas in a multiprocessor system, some 
processors could be serving the background queue while others are serving the 
foreground queue.

The solutions obtained enable us to evaluate and minimize a cost function which 
takes both holding and energy consumption costs into account. Among the 
numerical experiments that are carried out is one comparing the performance of 
the two-phase FB policy with that of FCFS and the pure LAS policy. The result 
shows an apparent (but explainable) violation of the optimality of LAS. Other 
experiments examine= the gains achieved by speed modulation in a single-server 
system and by capacity modulation in a multprocessor system. We also explore 
the possiblity of using the two-phase model in order to approximate 
three-phase models.

\subsection{Related literature}

The idea of scheduling jobs according to their attained service has been widely 
investigated in the literature. That area of study was opened up more than half a 
century ago with two seminal papers by Schrage \cite{schr}, and Coffman and 
Kleinrock \cite{coff} (see also \cite{klei}). Thresholds on the attained service were 
used to assign priorities to the waiting jobs, and service was given in quanta. The 
LAS policy emerged as a limiting case when the number of thresholds tends to 
infinity and the quantum size tends to 0. A good survey of subsequent 
developments can be found in Nuyens and Wierman \cite{nuy}. 

More recently, a large class of scheduling policies, including LAS, FIFO, SRPT and 
others, was analysed by Scully et al. \cite{scu1} in the context of a single server 
without speed modulation. That class is referred to as SOAP -- Schedule Ordered 
by Age-based Priority. Our policy is not in the SOAP class because it assigns 
priorities according to phase, rather than age, and the phase changes as the job 
progresses. In fact, we will show that scheduling based on the phase of 
service can lead to a lower expected response time than that provided by LAS. 

Speed scaling policies applied to several scheduling disciplines have also been widely studied, adopting different approaches to the trade-offs between energy saving and 
performance. For example, Yao et al. \cite{yao} analyse systems in which jobs
have deadlines and dynamic speed scaling is used. Bansal et al. \cite{ban}, examine 
the problem of minimizing the expected response time given a fixed energy budget. 

An M/M/1 queue with  occupancy-dependent server speed was analysed by George 
and Harrison \cite{geo}, with a view to minimizing average service costs (which 
may be interpreted as energy consumption costs). Performance was not included in 
the optimization. That study was later generalized by Wierman et al. \cite{wie} to 
M/G/1/PS queues. 

Those models do not allow the scheduling discipline to depend on job sizes. Such 
a dependence was included in Marin et al. \cite{mar}, where speed scaling was 
modelled in the context of a variant of the SRPT policy. In all these papers, the 
speed scaling strategy is energy-proportional, i.e. the power consumed by the 
processor depends on the speed at which it operates, and that in turn is determined 
by the number of jobs in the queue. That approach is commonly adopted in the 
literature (see Andrew and Wierman \cite{andr} and Bansal et al. \cite{ban9}). It is 
nearly optimal. 

Elahi et al. \cite{ela} have studied a threshold-based policy with a restricted form 
of speed modulation. Jobs that have received more than a certain amount of service 
are assigned a lower priority and speed modulation is applied to them only. Another 
example of server control involving the LAS policy is described by Lassila and Aalto
\cite{las}. In that work, LAS is combined with server sleeping states. The conclusion 
is that such an approach does not minimize either the linear combination of the 
expected response time and the energy consumption, or their product. 

The literature concerning scheduling policies in systems with multiple servers is 
not as large, but still quite extensive. Most of the tractable models in that area 
involve jobs of different types arriving in independent Poisson streams. 
Harchol-Balter et al. \cite{har} used phase-type distributions to approximate
various busy periods and recursively reduce the dimensionality of the model to 
one. The resulting QBD process is solved by matrix-analytic methods. The M/M/n  
model with two preemptive priority queues was studied by Mitrani and King 
\cite{mit}, by Gail at al.\cite{gai2} and by Kao and Narayanan \cite{kao1}. The 
case of non-preemptive priorities was examined by Gail at al.\cite{gai1} and by 
Kao and Wilson \cite{kao}. Kella and Yechiali \cite{kel} considered the special case 
of several priority queues with identical average service times for all job types. 

The optimal scheduling policy for a heavily loaded M/GI/n queue was established 
by Scully et al. \cite{scu2}. It turns out to be a version of the Gittins index policy. 
However, determining performance measures for that policy is intractable. 

We have not encountered in the literature an example exhibiting the features 
present in our model: multiple servers, non-exponential service times, two 
priority queues and capacity modulation.

\subsection{Structure of the paper}
In Section 2, we present the single-server model and in Section 3 we show
its exact solution. Section 4 discusses some special cases for which the solution 
can be obtained in closed form. The multiserver model is described and slved in 
Section 5 and in the Appendix. Numerical experiments and comparison with other 
disciplinesare discussed in Section 6. Section 7 concludes the paper. 

\section{The single-server model}

\begin{figure}[htb!]
	\begin{center}
		\includegraphics[width=0.4\textwidth]{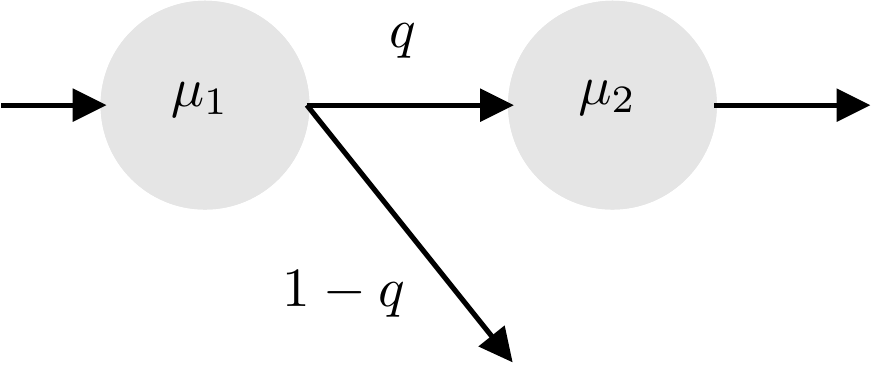}
	\end{center}
	\caption{Graphical representation of the 2-phase Coxian distribution.} 
	\label{fig:coxian}
\end{figure}

Jobs arrive in a Poisson stream at rate $\lambda$. Their lengths (measured in 
number of instructions) are i.i.d. random variables with a two-phase Coxian 
distribution (see \cite{cox}). Phase 1 is distributed exponentially with mean 
$1/\nu_1$. Phase 2 follows with probability $q$, and its length is distributed 
exponentially with mean $1/\nu_2$. That distribution is illustrated in Figure 
\ref{fig:coxian}.

\begin{figure}[tbp]
	\begin{center}
        \includegraphics[width=0.4\textwidth]{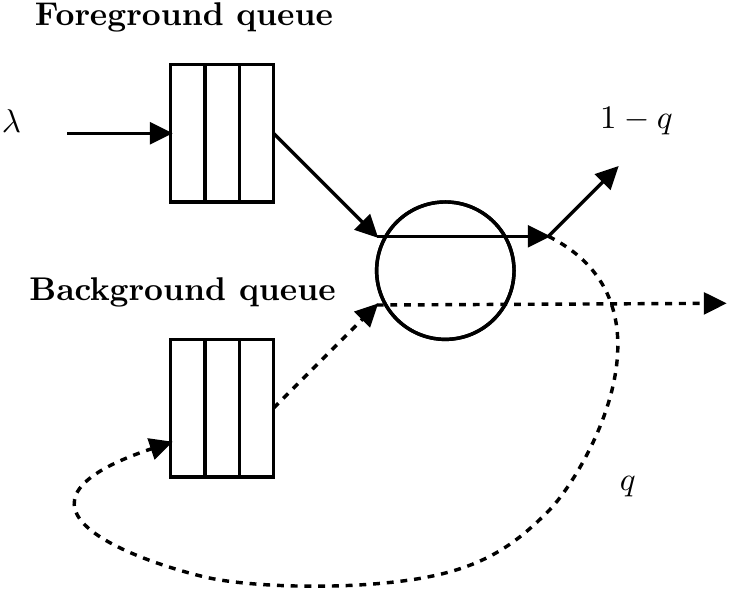}
	\end{center}
	\caption{Graphical representation of the Coxian service time distribution.}
	\label{fig:queues}
\end{figure}

On arrival, jobs join the foreground queue, where they execute phase 1. Upon 
completion, a job departs with probability $1-q$, and joins the background queue 
with probability $q$, in order to execute phase 2. However, the background queue 
is served only when the foreground one is empty. If a new job arrives during a 
background service, the latter is interrupted and the new job starts phase 1 in the 
foreground queue. This is a version of the Least Attained Service (LAS) policy 
where context is switched only at moments of arrival or phase completion. The 
queues and the flow of jobs are illustrated in Figure \ref{fig:queues}.

The processor speed (measured in instructions per second) can be controlled and 
depends on the total number of jobs in the system. There are $K+1$ possible speed l
evels, $s_0<s_1<\ldots <s_K$. The level is $s_n$ when the total number of jobs in 
the system is $n$, for $n=0,1,\ldots , K-1$, and it is $s_K$ when that number is greater 
than or equal to $K$. In other words, if the total number of jobs in the system is $n<K$, 
the service rates in the foreground and background queues are 
$\mu_{1,n}=\nu_1s_n$ and $\mu_{2,n}=\nu_2s_n$, respectively. For all $n\geq K$, 
those rates are $\mu_1=\mu_{1,K}=\nu_1s_K$ and 
$\mu_2=\mu_{2,K}=\nu_2s_K$, respectively. 

The system state is described by the pair of integers $(n_1,n_2)$, where $n_1$ is 
the number of jobs in the foreground queue and $n_1$ is the number of jobs in 
the background queue. Let $\pi_{i,j}$ be the steady-state probabilities of those 
states:
\begin{equation}
\pi_{i,j} = P(n_1=i,n_2=j) \;\;;\;\;i,j=0,1,\ldots\;.
\end{equation}

One might guess that the system is stable and steady-state exists when the 
processor, working at the highest speed level, can cope with the offered load:
\begin{equation} \label{erg}
\lambda (\frac{1}{\mu_1} + q\frac{1}{\mu_2}) < 1\;.
\end{equation}
This is indeed the case and will be established analytically.

The steady-state probabilities satisfy the following set of balance equations.

\noindent Case 1. $i>0,~i+j\geq K$ (processor serves the foreground queue at 
maximum speed):
\begin{equation} \label{iK}
(\lambda+\mu_1)\pi_{i,j} = \lambda \pi_{i-1,j} + \mu_1q\pi_{i+1,j-1}\delta(j>0) + \mu_1(1-q)\pi_{i+1,j} \;,\; 
\end{equation}
where $\delta(B)$ is 1 when the Boolean $B$ is true, 0 when false.

\noindent Case 2. $i>0,~i+j<K$ (processor serves the foreground queue at 
lower speed):
\begin{equation} \label{ij}
(\lambda+\mu_{1,i+j})\pi_{i,j} = \lambda \pi_{i-1,j} + 
\mu_{1,i+j}q\pi_{i+1,j-1}\delta(j>0) +  \mu_{1,i+j+1}(1-q)\pi_{i+1,j} \;.
\end{equation}

\noindent Case 3. $i=0,~j\geq K$ (processor serves the background queue at 
maximum speed):
\begin{equation} \label{0K}
(\lambda+\mu_2)\pi_{0,j} = \mu_2 \pi_{0,j+1} + \mu_1q\pi_{1,j-1} + 
\mu_1(1-q)\pi_{1,j} \;.
\end{equation}

\noindent Case 4. $i=0,~0<j<K$ (processor serves the background queue at 
lower speed):
\begin{equation} \label{0j}
(\lambda+\mu_{2,j})\pi_{0,j} = \mu_{2,j+1} \pi_{0,j+1}  + \mu_{1,j}q\pi_{1,j-1} + \mu_{1,j+1}(1-q)\pi_{1,j} \;.
\end{equation}

\noindent Case 5. $i=0,~j=0$ (processor is idle):
\begin{equation} \label{00}
\lambda\pi_{0,0} = \mu_{2,1} \pi_{0,1}  +  \mu_{1,1}(1-q)\pi_{1,0} \;.
\end{equation}

From the joint distribution $\pi_{i,j}$, it is possible to determine the marginal 
probabilities, $p_n$, that there is a total of $n$ jobs in the system ($n=0,1,\ldots$), 
and also the corresponding average number, $L$. When the processor runs at 
speed $s_k$, it consumes energy at a rate proportional to $s_k^\alpha$, where 
$\alpha$ is a constant which depends on the design of the processor; its value 
is usually between 1 and 3 (e.g., see \cite{wie}). To examine the 
trade-offs between holding costs and energy costs, we define a cost function, 
$C$, which is a linear combination of the two:
\begin{equation} \label{C}
C = c_1 L + c_2 \left [  \sum_{n=0}^{K-1}p_ns_n^\alpha + s_K^\alpha 
\sum_{n=K}^\infty p_n \right ]\;, 
\end{equation}
where $c_1$ and $c_2$ are coefficients reflecting the relative importance given 
to holding jobs in the system and energy consumption, respectively. 

The objective would be to choose the number and values of the speed levels 
$s_n$, so as to minimize the cost function $C$.

\section{Exact solution}

We start by concentrating on the system states $(i,j)$ where $i+j\geq K$, i.e. 
where the phase 1 and phase 2 completion rates are $\mu_1$ and $\mu_2$, 
respectively. In order to determine the probabilities corresponding to those 
states, it will be helpful to introduce the following generating functions.
\begin{equation} \label{g}
g_i(z) = \sum_{j=K-i}^\infty \pi_{i,j}z^j\;\;;\;\;i=0,1,\ldots K-1 \;,
\end{equation}
and
\begin{equation} \label{giK}
g_i(z) = \sum_{j=0}^\infty \pi_{i,j}z^j\;\;;\;\;i=K,K+1\ldots \;.
\end{equation}

The steady-state balance equations can now be transformed into relations 
between these functions, plus some of the unknown probabilities. 

Consider first the case $i=0$. Multiply equation (\ref{0K}) by $z^j$ and 
sum over all $j= K, K+1, \ldots\;$. This leads to
\begin{equation} \label{g00}
(\lambda+\mu_2)g_0(z) = \frac{\mu_2}{z}[g_0(z) -\pi_{0,K}z^K]+\mu_1(1-q)[g_1(z) -\pi_{1,K-1}z^{K-1}] 
\end{equation}
\[
{} + \mu_1q z g_1(z)\;.
\]
This can be rewritten as 
\begin{equation} \label{g0}
[\lambda+\mu_2\frac{z-1}{z}]g_0(z) = \mu_1(1-q+qz)g_1(z) 
\end{equation}
\[
{} - z^{K-1}[\mu_2\pi_{0,K}+\mu_1(1-q)\pi_{1,K-1}]\;.
\]
Similarly, for $i=1,2,\ldots ,K-1$, multiply (\ref{iK}) by $z^j$ and sum over 
all $j\geq K-i$. The resulting relation is
\begin{equation} \label{gi}
(\lambda+\mu_1)g_i(z) = \lambda g_{i-1}(z) + \mu_1(1-q+qz)g_{i+1}(z) 
\end{equation}
\[
{} + \lambda z^{K-i}\pi_{i-1,K-i} -\mu_1(1-q)z^{K-1-i}\pi_{i+1,K-1-i};. 
\]
When $i=K$, the last term in the right-hand side disappears:
\begin{equation} \label{gK}
(\lambda+\mu_1)g_K(z) = \lambda g_{K-1}(z) + \mu_1(1-q+qz)g_{K+1}(z) + \lambda \pi_{K-1,0}\;.
\end{equation}
For all $i>K$, the corresponding relations do not involve boundary probabilities 
and the sum ranges from 0 to $\infty$.
\begin{equation} \label{gKi}
(\lambda+\mu_1)g_i(z) = \lambda g_{i-1}(z) + \mu_1(1-q+qz)g_{i+1}(z) \;;\;i= K+1, K+2, \ldots\;.
\end{equation}
The next step is to combine all one-dimensional generating functions into a 
single two-dimensional function. For this purpose, we define
\begin{equation} \label{G0}
G(y,z) = \sum_{i=0}^\infty g_i(z) y^i\;.
\end{equation}
Multiply the equation for $g_i(z)$ by $y^i$ and sum over all $i=0,1,\ldots\;$. 
In order to facilitate the manipulations, add $\mu_1g_0(z)$ to both sides of 
(\ref{g0}). The equation that emerges is 
\begin{equation} \label{G}
(\lambda+\mu_1)G(y,z) = \lambda yG(y,z) +\frac{\mu_1(1-q+qz)}{y} 
[G(y,z)-g_0(z)] - \mu_2z^{K-1}\pi_{0,K} 
\end{equation}
\[
{} + [\mu_1 - \frac{\mu_2(z-1)}{z}]g_0(z) + 
\sum_{i=1}^Kz^{K-i}y^{i-1}[\lambda y\pi_{i-1,K-i}-\mu_1(1-q)\pi_{i,K-i}]\;.
\]
After multiplying both sides by $yz$ and rearranging terms, this can be 
rewritten as follows.
\begin{equation} \label{GG}
d(y,z)G(y,z) = a(y,z)g_0(z) + b(y,z)\;,
\end{equation}
where
\begin{equation} \label{d}
d(y,z)= z[\lambda y(1-y) - \mu_1(qz+1-q-y)]\;,
\end{equation}
\begin{equation} \label{a}
a(y,z)= \mu_1z(y-1+q-qz)+\mu_2y(1-z);,
\end{equation}
and
\begin{equation} \label{b}
b(y,z)= - \mu_2yz^K\pi_{0,K}  + \sum_{j=1}^Kz^{K+1-j}y^j
[\lambda y \pi_{j-1,K-j}-\mu_1(1-q)\pi_{j,K-j}]\;.
\end{equation}

Thus the two-dimensional generating function $G(y,z)$ is expressed in 
terms of the one-dimensional function $g_0(z)$ and the probabilities that 
appear in (\ref{b}). In order to determine $g_0(z)$, note that for every value 
of $z$ in the interval (0,1), the coefficient $d(y,z)$ satisfies 
$d(0,z)<0$, $d(1,z)>0$ and $d(\infty,z)<0$. The polynomial $d(y,z)$ is 
quadratic in $y$, hence for every such $z$ it has exactly two real zeros, 
$y_1(z)$ and $y_2(z)$, in the intervals (0,1) and $(1,\infty)$, respectively. 
After dividing the numerator and denominator in the expression for $y_1(z)$ 
by $\mu_1$, and rearranging terms, that expression can be written as 
\begin{equation} \label{yz}
y_1(z)= \frac{1+\rho_1-\sqrt{(1-\rho_1)^2-4\rho_1q(z-1)}}{2\rho_1} \;.
\end{equation}
The minus sign in front of the square root is taken because $y_1(z)$ is the 
smaller of the two zeros. 

We shall also need the first and higher order derivatives of $y_1(z)$. These 
are given by 
\begin{equation} \label{yz1}
y_1^\prime(z)= \frac{q}{\sqrt{(1-\rho_1)^2-4\rho_1q(z-1)}} \;,
\end{equation}
and
\begin{equation} \label{yzn}
y_1^{(n)}(z)= y_1^{(n-1)}(z)\frac{2(2n-3)\rho_1 q}{(1-\rho_1)^2-4\rho_1q(z-1)} \;\;;\;\; n=2,3,\ldots \;.
\end{equation}
In particular, we have
\begin{equation} \label{y1}
y_1(1) = 1\;\;;\;\; y_1^\prime (1) = \frac{q}{1-\rho_1}\;\;;\;\; y_1^{\prime\prime}(1) = 
\frac{2\rho_1 q^2}{(1-\rho_1)^3}\;. 
\end{equation}

Since the generating function $G(y,z)$ is finite when $0\leq y \leq 1$ and 
$0\leq z \leq 1$, the right-hand side of (\ref{GG}) must vanish when 
$y=y_1(z)$. This provides an expression for $g_0(z)$ in terms of the 
probabilities in (\ref{b}).
\begin{equation} \label{g0z}
a(y_1(z),z)g_0(z) = - b(y_1(z),z)\;.
\end{equation}
In fact, that expression can be simplified a little. From $d(y_1(z),z)=0$ it 
follows that 
\begin{equation} \label{mz}
\mu_1z(qz+1-q-y_1(z)) = \lambda zy_1(z)(1-y_1(z))\;.
\end{equation}
Substituting this into (\ref{a}) yields
\begin{equation} \label{a1}
a(y_1(z),z)= y_1(z)[\mu_2(1-z) - \lambda z(1-y_1(z))];.
\end{equation}
This allows us to cancel a factor of $y_1(z)$ from both sides of (\ref{g0z}), 
reducing that expression to
\begin{equation} \label{gz}
g_0(z) = \frac{\mu_2z^K\pi_{0,K}  - \sum_{j=1}^Kz^{K+1-j}y_1(z)^{j-1}
[\lambda y_1(z)\pi_{j-1,K-j}-\mu_1(1-q)\pi_{j,K-j}]}{\mu_2(1-z) - \lambda z(1-y_1(z))}\;.
\end{equation}

We are now left with $(K+1)(K+2)/2$ unknown probabilities, 
including the ones involved in (\ref{b}) and those not included in 
the definitions of the generating functions. These unknowns are 
$\pi_{i,j}$, for $i+j\leq K$. The balance equations (\ref{ij}), (\ref{0j}) and 
(\ref{00}) provide $K(K+1)/2$ relations among them. Another $K$ equations 
are obtained as follows. Divide both sides of (\ref{gz}) by $z^K$ and 
let $z\rightarrow 0$. Remembering that the first term in the expansion of 
$g_0(z)$ is $\pi_{0,K} z^K$, we conclude that when $z\rightarrow 0$,
\begin{equation} \label{lim}
\frac{\sum_{j=1}^K z^{K-j}y_1(z)^{j-1} [\lambda y_1(z)\pi_{j-1,K-j}-\mu_1(1-q)\pi_{j,K-j}]}{z^{K-1}} 
\rightarrow 0\;.
\end{equation}
For that limit to hold, the first $K$ terms in the Maclaurin series expansion 
of the numerator in (\ref{lim}) must be 0. In other words, the numerator itself, 
together with its first $K-1$ derivatives, must vanish at $z=0$. This provides 
$K$ additional equations for the unknown probabilities. The 
derivatives of $y_1(z)$ at $z=0$ are given by (\ref{yz1}) and (\ref{yzn}).

The final equation that we need is provided by the normalizing condition
\begin{equation} \label{norm}
G(1,1) + \sum_{i=0}^{K-1}\sum_{j=0}^{K-1-i} \pi_{i,j} = 1\;.
\end{equation}

To find an expression for $G(1,1)$, set $y=z$ in (\ref{G}). We have 
\begin{equation} \label{dd}
d(z,z)= z(1-z)[\lambda z - \mu_1(1-q)]\;,
\end{equation}
\begin{equation} \label{aa}
a(z,z)= z(1-z)[\mu_2-\mu_1(1-q)]\;,
\end{equation}
and 
\begin{equation} \label{bb}
b(z,z)= z^{K+1} [- \mu_2\pi_{0,K}  + \sum_{j=1}^K \lambda z\pi_{j-1,K-j}-
\mu_1(1-q)\pi_{j,K-j}  ]\;.
\end{equation}

Since both $d(z,z)$ and $a(z,z)$ are 0 at $z=1$, so is $b(z,z)$. In other 
words,
\begin{equation} \label{mm}
\lambda\sum_{j=1}^K \pi_{j-1,K-j} = \mu_2\pi_{0,K}  + 
\mu_1(1-q)\sum_{j=1}^K \pi_{j,K-j}\;.
\end{equation}
The sum in the left-hand side of this equation is the probability, $p_{K-1}$, 
that the total number of jobs present is $K-1$. The equation expresses the 
balance of flow between states with $K-1$ 
or fewer jobs in the system, and those with $K$ or more. 

Using (\ref{mm}), expression (\ref{bb}) can be rewritten as 
\begin{equation} \label{bbb}
b(z,z)= - \lambda z^{K+1}(1-z) p_{K-1}\;.
\end{equation}

Substituting (\ref{dd}), (\ref{aa}) and (\ref{bbb}) into (\ref{GG}) we obtain
\begin{equation} \label{GGG}
G(z,z)= \frac{[\mu_1(1-q)-\mu_2]g_0(z) + \lambda z^K  p_{K-1}}{\mu_1(1-q) 
	- \lambda z}\;.
\end{equation}
This could also have been derived directly, as the generating function of 
the probabilities $p_K$, $p_{K+1}$, $\ldots$ . 

To complete the expression of $G(1,1)$ in terms of $\pi_{i,j}$, we need an 
expression for $g_0(1)$. This is provided by (\ref{gz}), after an application 
of L'Hospital's rule. The numerator and denominator in the right-hand side 
of (\ref{gz}) are both 0 at $z=1$, which means that derivatives 
must be taken at that point.

Having computed the unknown probabilities, one can evaluate the cost 
function (\ref{C}). The average total number of jobs in the system, $L$, is 
given by
\begin{equation} \label{L}
L = \sum_{i=0}^{K-1}\sum_{j=0}^{K-1-i} (i+j) \pi_{i,j}  + G^\prime (1,1) \;.
\end{equation}
The derivative of $G(z,z)$ at $z=1$ requires $g_0^\prime (1)$. The latter is 
again given by (\ref{gz}), after a double application of L'Hospital's rule.

\section{Special cases}

The simplest non-trivial special case for this model is $K=1$. The processor 
has two possible speeds: $s_0$ when idle and $s_1$ when there is at least 
one job present. The service rates in the foreground and background queues 
are $\mu_1$ and $\mu_2$ respectively, whenever those queues are served. 
The only state probability that is not included in the definitions of $g_i(z)$, 
(\ref{g}) and (\ref{giK}), is $\pi_{0,0}$. Hence, $\pi_{0,0} + G(y,z)$ is the full 
generating function of the joint distribution of the two queues. Also, 
$\pi_{0,0} + g_0(z)$ is the generating function 
of the state probabilities where the foreground queue is empty.

In this special case, it is intuitively clear that the foreground queue behaves 
like an M/M/1 queue with offered load $\rho_1 = \lambda /\mu_1$. To prove that, 
set $z=1$ in (\ref{GG}), for $K=1$:
\begin{equation} \label{Gy}
(1-y)(\mu_1-\lambda y)G(y,1)= \mu_1(1-y)g_0(1) +y [\mu_2\pi_{0,1} - 
\lambda y \pi_{0,0} + \mu_1(1-q)\pi_{1,0}]\;.
\end{equation}
Using the balance equation (\ref{00}) and cancelling the factor $1-y$, 
(\ref{Gy}) can be rewritten as 
\begin{equation} \label{Gy1}
G(y,1)= \frac{\mu_1g_0(1) + \lambda y\pi_{0,0}}{\mu_1-\lambda y} \;,
\end{equation}
or
\begin{equation} \label{Gy2}
G(y,1)= \frac{g_0(1) + \rho_1 y\pi_{0,0}}{1-\rho_1 y} \;.
\end{equation}

Adding $\pi_{0,0}$ to both sides produces 
\begin{equation} \label{Gy3}
\pi_{0,0} + G(y,1) = \frac{g_0(1) + \pi_{0,0}}{1-\rho_1 y} \;.
\end{equation}

This is the generating function for the geometric distribution of an M/M/1 
queue. The normalizing condition yields
\begin{equation} \label{nor}
g_0(1) + \pi_{0,0} = 1-\rho_1 \;.
\end{equation} 
As expected, the average number of jobs in the foreground queue, $L_1$, is
\begin{equation} \label{L1}
L_1 = \frac{\rho_1}{1-\rho_1} \;.
\end{equation} 
Clearly, the condition $\rho_1<1$ is necessary for the stability of the foreground 
queue. However, it is not sufficient for the ergodicity of the model because the 
background queue may still be saturated. 

Expression (\ref{gz}) for $g_0(z)$, together with (\ref{00}), now has the form
\begin{equation} \label{g0z1}
g_0(z) = \frac{\lambda z\pi_{0,0} (1-y_1(z))}{\mu_2(1-z) - \lambda z(1-y_1(z))}\;.
\end{equation}
It is convenient to rewrite this in terms of the function $u(z)=(1-y_1(z))/(1-z)$:
\begin{equation} \label{g0z2}
g_0(z) = \frac{\rho_2 z\pi_{0,0} u(z)}{1 - \rho_2z u(z)}\;,
\end{equation}
where $\rho_2 = \lambda /\mu_2$. It is not difficult to see that $u(z)$ satisfies
\begin{equation} \label{u}
u(1) = y_1^\prime(1)\;\;;\;\; u^\prime (1) = \frac{y_1^{\prime\prime}(1)}{2}\;.
\end{equation}
The values of $y_1^\prime(1)$ and $y_1^{\prime\prime}(1)$ are given by (\ref{y1}).

Substituting (\ref{u}) and (\ref{y1}) into (\ref{g0z2}) gives
\begin{equation} \label{g0z3}
g_0(1) = \frac{\rho_2 q\pi_{0,0}}{1 - \rho_1 -\rho_2q}\;,
\end{equation}
which, together with the normalization (\ref{nor}) determines $\pi_{0,0}$ 
and $g_0(1)$:
\begin{equation} \label{pg1}
\pi_{0,0} = 1-\rho_1-\rho_2q\;\;;\;\; g_0(1) = \rho_2q\;.
\end{equation}
This result shows that a normalizable positive solution for the probabilities 
$\pi_{i,j}$ exists when $\rho_1+\rho_2q <1$. That inequality, which coincides 
with (\ref{erg}), is therefore the ergodicity condition for the $K=1$ model.

In order to determine the average total number of jobs in the system, $L$, one could 
differentiate (\ref{GGG}) at $z=1$. Alternatively, since we already know $L_1$, it is 
simpler to find the average number of jobs in the background queue, $L_2$. Setting 
$y=1$ in (\ref{GG}), using the balance equation (\ref{00}) and cancelling the factor 
$1-z$, we obtain
\begin{equation} \label{G1z}
G(1,z) = g_0(z)(1 + \frac{\mu_2}{\mu_1qz}) = g_0(z)(1 + \frac{\rho_1}{\rho_2qz})\;.
\end{equation}
The average $L_2$ is the derivative of the right-hand side at $z=1$. To find 
$g_0^\prime(1)$, differentiate (\ref{g0z2}) and use (\ref{u}). This yields
\begin{equation} \label{g01}
g_0^\prime (1) = \frac{\rho_2q}{1-\rho_1 -\rho_2q}[1-\rho_1+\frac{\rho_1q}{1-\rho_1}]\;,
\end{equation}
and
\begin{equation} \label{L2}
L_2 = \frac{\rho_1 +\rho_2q}{1-\rho_1 -\rho_2q}[1-\rho_1+
\frac{\rho_1q}{1-\rho_1}]-\rho_1\;.
\end{equation}
Thus the complete solution of the $K=1$ model is obtained in closed form. In this case, 
the idling speed of the processor only affects the energy costs, not the holding ones. 
Therefore, in order to minimize the cost function (\ref{C}), $s_0$ should be set as low 
as possible.

Another related special case worth examining is the one where $K$ is arbitrary, 
but $s_n =0$ for $n<K$. That is, the processor works only when there are $K$ 
or more jobs present. Once the background queue reaches size $K-1$, it cannot 
drop down below that level because it can be served only when the foreground 
queue is empty and then the processor stops working. Hence, all states $(i,j)$, 
such that $j<K-1$, are transient and their long-term probabilities are zero.

The main interest of this model is that it defines the stability region for a given 
top speed $s_K$. Indeed, if the system is stable when the lower service rates 
are zero (in the sense that the recurrent states have normalizable probabilities), 
then it would be stable when those service rates are greater than zero. 

The new generating functions are almost identical to the ones for the case $K=1$. 
The role of  $\pi_{0,0}$ is now played by $\pi_{0,K-1}$. Since $\pi_{i,j}=0$ for 
$j<K-1$, the balance equation for $\pi_{0,K-1}$ is similar to (\ref{00}):
\begin{equation} \label{000}
\lambda\pi_{0,K-1} = \mu_2\pi_{0,K}  + \mu_1(1-q)\pi_{1,K-1} \;.
\end{equation}
Equations (\ref{Gy1}) - (\ref{nor}) are valid, with $\pi_{0,0}$ replaced by 
$\pi_{0,K-1}$. The foreground queue again behaves like an M/M/1 queue with 
offered load $\rho_1$, and its average size, $L_1$, is given by (\ref{L1}). 

The expression for the generating function $g_0(z)$ is similar to (\ref{g0z2}), 
except that the factor $z$ is now $z^K$:
\begin{equation} \label{g0z4}
g_0(z) = \frac{\rho_2 z^K\pi_{0,K-1} u(z)}{1 - \rho_2z u(z)}\;,
\end{equation}
The result (\ref{pg1}) becomes
\begin{equation} \label{pg2}
\pi_{0,K-1} = 1-\rho_1-\rho_2q\;\;;\;\; g_0(1) = \rho_2q\;.
\end{equation}

We conclude that the stability condition (\ref{erg}) holds in any system where 
the arrival rate is  $\lambda$, the probability of phase 2 is $q$ and the top 
service rates are $\mu_1$ and $\mu_2$.

Differentiating (\ref{G1z}) at $z=1$ yields the average size of the background 
queue:
\begin{equation} \label{L2K}
L_2 = (K-1)(\rho_1 +\rho_2q)  + \frac{\rho_1 +\rho_2q}{1-\rho_1 -\rho_2q} 
[1-\rho_1+\frac{\rho_1q}{1-\rho_1}]-\rho_1\;.
\end{equation}

Intuitively, this model minimizes the energy costs and maximizes the holding 
costs, given the top speed of the processor. 

\section{Multiple servers} \label{sec:mult}

Instead of a single speed-modulated server, we now consider a system with 
$m$ identical parallel servers. We start by assuming that $m$ is fixed, but later 
we shall allow that number to be controlled dynamically, for purposes of energy 
conservation. In all other respects, the model is the same as before. The crucial 
new feature of the resulting two-dimensional Markov process is that, if the 
number of jobs in the foreground queue is less than $m$, some servers may 
be serving the background queue while others are serving the foreground one. 
This changes the analysis substantially, although the general apprach based 
on generating functions remains the same.

The offered load in the foreground queue is $\rho_1=\lambda /\mu_1$; in the 
background queue it is $\rho_2=\lambda q/\mu_2$. The whole system is stable 
and steady-state exists when the total offered load is lower than 
the number of available servers:
\begin{equation} \label{ergm}
	\rho_1 + \rho_2 < m\;.
\end{equation}
This intuitive condition can also be established analytically. 

Denote again by $\pi_{i,j}$ the steady-state probability that there are $i$ jobs 
in the foreground queue and $j$ jobs in the background queue. There is no 
difficulty in computing the marginal distribution and the average number of 
jobs in the foreground queue, $L_1$, since that queue behaves like an $M/M/m$ 
queue with parameters $\lambda$ and $\mu_1$. However, in order to find the 
average number in the background queue, $L_2$, it is necessary to determine 
the joint distribution, $\pi_{i,j}$, of the two queues.

We shall write separately the balance equations for $i<m$, when $i$ servers are 
allocated to the foreground queue and $m-i$ servers are available for the 
background queue, and for $i\geq m$, when the service rates at the foreground 
and background queues are $m\mu_1$ and 0, respectively.
\[
[\lambda +i\mu_1 +\min(j,m-i)\mu_2]\pi_{i,j} = \lambda \pi_{i-1,j} +
(i+1)\mu_1 [(1-q)\pi_{i+1,j}  + q\pi_{i+1,j-1}] 
\]
\begin{equation} \label{bij} 
	{} + \min(j+1,m-i)\mu_2\pi_{i,j+1} \;\;;\;\;i=0,1,\ldots ,m-1\;,\;j\geq 0\;,
\end{equation}
\begin{equation} \label{bmj}
	(\lambda +m\mu_1)\pi_{i,j} = \lambda \pi_{i-1,j} +m\mu_1 [(1-q)\pi_{i+1,j}  
	+ q\pi_{i+1,j-1}] \;;\;i\geq m\;,\;j\geq 0\;.
\end{equation}
In both cases, any probability with a negative index is 0 by definition. 

To solve these equations, introduce the generating functions
\begin{equation} \label{gim}
	g_i(z) = \sum_{j=0}^\infty \pi_{i,j} z^j\;;\; i=0,1,\ldots\;.
\end{equation}

Focusing first on the case where the background queue service rate is 0, 
we multiply the $j$th equation in (\ref{bmj}) by $z^j$ and sum over all $j$.  
This leads to relations similar to (\ref{gKi}):
\begin{equation} \label{gmz}
(\lambda +m\mu_1)g_i(z) = \lambda g_{i-1}(z) + 
m\mu_1 (1-q+qz) g_{i+1}(z)\;;\;i\geq m\;.
\end{equation}
These can be combined into a single equation for the two-dimensional generating 
function, $g(y,z)$, defined as 
\begin{equation} \label{gyzm}
	g(y,z) = \sum_{i=m}^\infty g_i(z) y^{i-m}\;.
\end{equation}
Multiplying (\ref{gmz}) by $y^{i-m}$ and summing over all $i\geq m$ expresses 
$g(y,z)$ in terms of $g_{m-1}(z)$ and $g_{m}(z)$:
\begin{equation} \label{gyz}
	d(y,z)g(y,z) = \lambda yg_{m-1}(z) - m\mu_1(1-q+qz)g_{m}(z)\;,
\end{equation}
where
\[
d(y,z) = \lambda y(1-y)+m\mu_1(y-1+q-qz)\;.
\]
Now note that for every value of $z$ in the interval (0,1), the coefficient $d(y,z)$, 
which is quadratic in $y$, is negative at $y=0$, positive at $y=1$ and negative at 
$y=\infty$. Therefore, $d(y,z)$ has exactly two real zeros, $y_1(z)$ and $y_2(z)$, 
such that $0<y_1(z)\leq 1<y_2(z)<\infty$. Since $g(y,z)$ is finite on the closed 
interval $[0,1]$, the right-hand side of (\ref{gyz}) must vanish at $y=y_1(z)$. 
This implies a relation between $g_{m-1}(z)$ and $g_{m}(z)$:
\begin{equation} \label{gm}
	\lambda y_1(z)g_{m-1}(z) = m\mu_1(1-q+qz)g_{m}(z)\;.
\end{equation}

Expressions (\ref{gyz}) and (\ref{gm}) can be simplified further by using the 
properties of the quadratic:
\begin{equation} \label{ayz}
	d(y,z) = \lambda [y-y_1(z)][y_2(z)-y]\;,
\end{equation}
and
\begin{equation} \label{y12}
	\lambda y_1(z)y_2(z) = m\mu_1(1-q+qz)\;.
\end{equation}
Substituting (\ref{y12}) into (\ref{gm}) yields
\begin{equation} \label{gm1}
	g_{m-1}(z) = y_2(z)g_{m}(z)\;.
\end{equation}
Then, substituting (\ref{ayz}), (\ref{gm}) and (\ref{gm1}) into (\ref{gyz}), the 
latter becomes
\begin{equation} \label{gyz1}
	g(y,z) = \frac{g_{m-1}(z)}{y_2(z) - y}\;.
\end{equation}

The generating function equations for $i<m$ are more complicated because they 
include terms involving $\mu_2$, and depend on probabilities $\pi_{i,j}$ for 
$j<(m-i)$. After some manipulations we obtain 
\[
[\lambda z+i\mu_1 z+(m-i)\mu_2(z-1)]g_i(z) = \lambda z g_{i-1}(z) 
\]
\begin{equation} \label{giz1}
	{} + (i+1)\mu_1 z(1-q+qz) g_{i+1}(z) + b_i(z) \;\;;\;\; i=0,1,\ldots , m-1\;,
\end{equation}
where $g_{-1}(z)=0$ by definition and
\[
b_i(z) = \mu_2 (z-1) \sum_{j=0}^{m-i-1} (m-i-j)\pi_{i,j} z^j \;;\;i=0,1,\ldots ,m-1\;.
\]

The term $m\mu_1 z(1-q+qz) g_m(z)$ which appears in the right-hand side of 
(\ref{giz1}) for $i=m-1$ can be replaced by $\lambda z y_1(z)g_{m-1}(z)$, 
according to (\ref{y12}) and (\ref{gm1}). Then (\ref{giz1}) become a set of 
simultaneous linear equations for the generating functions $g_0(z)$, $g_1(z)$, 
$\ldots$, $g_{m-1}(z)$. We shall write these in matrix and vector form as follows. 
\begin{equation} \label{Az}
	A(z)\mathbf{g}(z) = \mathbf{b}(z)\;,
\end{equation}
where $\mathbf{g}(z)$ is the column vector $[g_0(z),g_1(z),\ldots,g_{m-1}(z)]$,
$\mathbf{b}(z)$ is the column vector $[b_0(z),b_1(z),\ldots,b_{m-1}(z)]$, and
$A(z)$ is a tri-diagonal matrix
\[
A(z) = \left [ \begin{array}{cccccc} a_0(z) & -\alpha_1(z) & & & & \\
	-\lambda z & a_1(z) & -\alpha_2(z) & & & \\
	& -\lambda z & a_2(z) & -\alpha_3(z) & & \\
	& & & \ddots & & \\
	& & & -\lambda z & a_{m-2}(z) & -\alpha_{m-1}(z) \\
	& & & & -\lambda z & a_{m-1}(z)
\end{array} \right ] \;.
\]
The diagonal elements of $A(z)$ are
\[
a_i(z) = \lambda z +i\mu_1 z +(m-i)\mu_2(z-1)\;\;;\;\;i=0,1,\ldots ,m-2\;,
\]
and 
\[
a_{m-1}(z) = \lambda z[1-y_1(z)] +(m-1)mu_1 z +\mu_2(z-1) \;,
\]
while the upper diagonal elements are
\[
\alpha_i(z) = i\mu_1 z(1-q+qz) \;;\;i=1,2,\ldots ,m-1\;.
\]

The solution of (\ref{Az}) is given by
\begin{equation} \label{gs}
	g_i(z) = \frac{D_i(z)}{D(z)}\;;\;i=0,1,\ldots,m-1\;,
\end{equation}
where $D(z)$ is the determinant of $A(z)$ and $D_i(z)$ is the
determinant of the matrix obtained from $A(z)$ by replacing its $i+1$'st
column with the column vector $\mathbf{b}(z)$ (the columns are numbered 
1 to $m$, rather than 0 to $m-1$).

Thus all generating functions are determined in terms of the $m(m+1)/2$ 
unknown probabilities that appear in the elements of $\mathbf{b}(z)$: 
$\pi_{i,j}$, for $i<m$ and $j<m-i$. The balance equations (\ref{bij}), for 
$i<m-1$ and $j<m-i-1$, offer $(m-1)m/2$ relations between them. That 
is $m$ fewer than the number of unknowns. 

Another equation comes from the normalizing condition, requiring 
that the sum of $\pi_{i,j}$, for all $i$ and $j$, is 1. This can be expressed as
\begin{equation} \label{normm}
	\sum_{i=0}^{m-1}g_i(1) + g(1,1) = 1\;.
\end{equation}

The remaining $m-1$ equations needed in order to determine the unknown 
probabilities are provided by the  following result.

{\bf Lemma}. {\em When the stability condition (\ref{ergm}) holds, the 
determinant $D(z)$ has exactly $m-1$ distinct real zeros, $z_1$, $z_2$, 
$\ldots$, $z_{m-1}$, in the open interval (0,1).}

The proof of this Lemma is in the Appendix.

Consider one of the generating functions, say $g_0(z)$. Since it is finite 
on the interval (0,1), the numerator in the right-hand side of (\ref{gs}), 
$D_0(z)$, must vanish at each of the points $z_1$, $z_2$, $\ldots$, 
$z_{m-1}$, yielding $m-1$ equations. Using a different generating 
function would not provide new information.

One way of converting the normalizing equation (\ref{normm}) into an 
explicit relation for the unknown probabilities is by invoking the known 
distribution of the $M/M/m$ queue. The marginal probability that the 
foreground queue is empty is given by
\begin{equation} \label{g0m}
	g_0(1) = \frac{D_0(1)}{D(1)} = 
	\left [ \sum_{k=0}^{m-1} \frac{\rho_1^k}{k!} + 
	\frac{m\rho_1^m}{(m-\rho_1 )m!} \right ]^{-1}\;.
\end{equation}

Both $D_0(z)$ and $D(z)$ are 0 at $z=1$. One would have to resolve the 
indeterminacy and then evaluate the determinants, which is quite a 
laborious process. Fortunately, there is an argument that produces the 
desired equation directly, and in a simple form: 
\begin{equation} \label{s}
	\sum_{n=0}^{m-1} (m-n)\sum_{i=0}^{n} \pi_{i,n-i}= m-\rho_1-\rho_2\;.
\end{equation}
Indeed, the left-hand side of (\ref{s}) contains an expression for the average 
number of idle servers. On the other hand, Little's result implies that the 
average number of servers allocated to the foreground queue is $\rho_1$, 
and the average number of servers allocated to the background queue is 
$\rho_2$. Hence, the two sides evaluate the same quantity.

The above equation confirms the necessity of (\ref{ergm}) for stability. If 
a normalizeable positive solution to the balance equation exists, the right-hand 
side of (\ref{s}) must be positive. The sufficiency of the condition follows 
from the Lemma, since it enables the balance equations to be solved.

Having computed the unknown probabilities, the average number of jobs in 
the background queue is obtained from
\begin{equation} \label{L2m}
	L_2 =\sum_{i=0}^{m-1} g_i^\prime(1) + \frac{\partial}{\partial z} g(1,1)\;.
\end{equation}

The derivatives $g_i^\prime(1)$ can be computed either by applying the rule 
for differentiating a determinant, or more simply by using the definition of a 
derivative:
\begin{equation} \label{dgi}
	g_i^\prime(1) \approx \frac{g_i(1)-g_i(\delta)}{1-\delta}\;,
\end{equation}
for some value of $\delta$ suitably close to 1. In fact, it is enough to do that 
for $g_0^\prime(1)$. The other derivatives can be obtained by differentiating 
the recurrences (\ref{giz1}) at $z=1$ and using the known values of $g_i(1)$.

The last term in the right-hand side of (\ref{L2m}), given by (\ref{gyz1}), 
involves the larger zero, $y_2(z)$, of the coefficient $d(y,z)$ 
appearing in (\ref{gyz}). That function can be shown to satisfy
\begin{equation} \label{y2}
	y_2(1) = \frac{m\mu_1}{\lambda_1}\;\;;\;\; y_2^\prime(1)
	= - \frac{m\mu_1 q}{m\mu_1 -\lambda} \;.
\end{equation}

All elements of an algorithm for computing the desired performance measures 
are now in place. The following steps are carried out.
\begin{enumerate}	
	\item Compute the $m-1$ zeros of $D(z)$ in the interval (0,1).
	\item Assemble and solve the set of $m(m+1)/2$ equations for the unknown 
	probabilities.
	\item Compute $L_2$ from (\ref{L2m}).
	\item Compute $L_1$ from the $M/M/m$ result.
\end{enumerate}

\subsection{Energy saving}

Consider now the problem of controlling the number of servers dynamically, 
so as to optimize the trade-off between performance and energy consumption. 
Since an operative processor consumes energy when it is idle as well as when 
it is busy, it may be desirable to switch processors off when they are idle, and 
even when there are jobs that they could serve. One would then be trading 
energy costs against performance. 

Assume that processors can be switched off and on instantaneously, or at 
least fast enough compared to the interarrival and service times. Consider the 
following simple control policy which depends on a threshold parameter, 
$K = 0,1,\ldots , m-1$. Whenever the total number of jobs in te system drops 
down to $K$, all $m$ processors are switched off; at the next arrival instant 
they are switched back on again. Thus, if $K=0$, the processors are switched 
off when the system is empty; as far as performance is concerned, that is the 
case analysed in the previous section. If $K=m-1$, the servers are operative 
only when there are $m$ or more jobs present. 

The stability condition remains unchanged. It is not affected by the value of 
$K$, since in all cases servers are switched off only when there are fewer than 
$m$ jobs present.

The trade-off between holding costs and energy costs may be evaluated by 
defining a cost function, $C$, which is a linear combination of the two:
\begin{equation} \label{Cm}
	C = c_1 L + c_2 U\;, 
\end{equation}
where $L=L_1+L_2$ is the average total number of jobs present and $U$ is the 
average number of operative servers; $c_1$ and $c_2$ are coefficients 
reflecting the relative importance given to holding jobs in the system and 
energy consumption, respectively. 

To evaluate the cost function it is necessary to determine the joint distribution 
of the two queues, $\pi_{i,j}$. Note that once the total number of jobs reaches 
$K$, it can never drop back below that level because the servers stop working. 
Hence, all states $(i,j)$ such that $i+j<K$ are transient and their long-term 
probabilities are 0. Because of this, the marginal distribution of the foreground 
queue is no longer given by the $M/M/m$ result. The computations of $L_1$ 
and $L_2$ must be modified appropriately. 

The balance equations corresponding to states $(i,j)$ where the servers do not 
work are now 
\begin{equation} \label{pijK}
	\lambda \pi_{i,j} = (i+1)(1-q)\mu_1 \pi_{i+1,j} + (j+1)\mu_2 \pi_{i,j+1}\;;\; i+j=K<m\;. 
\end{equation}

All other balance equations, for $i+j>K$, are as before.

Since the number of operative processors is either 0 or $m$, the average $U$ 
is given by 
\begin{equation} \label{U}
	U = m \left [ 1 - \sum_{i=0}^K \pi_{i,k-i} \right ]\;;\;K=0,1,\ldots , m-1 \;. 
\end{equation}

The objective of the evaluation would be to choose the threshold $K$ so as to 
minimize (\ref{Cm}). Intuitively, setting $K=0$ would minimize $L$ and maximize $U$, 
while $K=m-1$ would maximize $L$ and minimize $U$. 

The analysis of the threshold policy proceeds along similar lines to the one 
described in the previous section. In particular, the generating functions and 
the relations between them have the same form as before. Only the terms 
$b_i(z)$ in equations (\ref{giz1}) change, and only for $i\leq K$. These 
now take into account the fact that $\pi_{i,j}=0$ when $i+j<K$, and also that 
all servers are off when $i+j=K$. The new expressions for $b_i(z)$ are
\[
b_i(z) = \{[i\mu_1z + (m-i)\mu_2 (z-1)] \pi_{i,K-i} - 
(i+1)\mu_1 (1-q+qz)\pi_{i+1,K-i-1} \} z^{K-i}
\]
\begin{equation} \label{bizK}
	{} + \mu_2 (z-1) \sum_{j=K-i+1}^{m-i-1} (m-i-j)\pi_{i,j} z^j \;\;;\;\;i=0,1,\ldots ,K\;.
\end{equation}

The number of unknown probabilities is $(m+K+1)(m-K)/2$, and the number of 
balance equations available for them is $(m+K)(m-K-1)/2$. The shortfall of $m$ 
equations is filled by the Lemma, which provides $m-1$ equations, and by the 
normalizing condition, which can be written as
\begin{equation} \label{sK}
	m \sum_{i=0}^{K} \pi_{i,K-i} + \sum_{n=K+1}^{m-1} 
	(m-n)\sum_{i=0}^{n} \pi_{i,n-i} = m-\rho_1-\rho_2\;.
\end{equation}
The two sides of this equation are equal to the average number of servers that 
are not serving jobs, i.e. are either idle or stopped. 

Although the full $M/M/m$ result no longer applies to the 
foreground queue, it is still true that the tail of its 
distribution, for $i\geq m$, is geometric with parameter 
$r =\lambda /(m\mu_1)$. Hence, the value of 
$L_1$ can be obtained from
\begin{equation} \label{L1K}
	L_1 =\sum_{i=0}^{m-1} i g_i (1) + g_m (1) 
	\frac{m(1-r)+r}{(1-r)^2}\;.
\end{equation}
The average number of jobs in the background queue is determined according 
to (\ref{L2m}), as before.

The cost function can now be computed for all values of the threshold parameter.

Clearly, there are other control policies that might be considered. For example, 
servers could be switched off individually, or in groups of varying size, depending 
on the number of jobs in the system. The cost function may also be generalized. 
In addition to the energy cost of running an operative server, there may be unit 
costs incurred each time a server is switched off and on. The analysis described 
here can be adapted to all such models, as long as all servers become operative 
whenever there are $m$ or more jobs present. Then the condition for stability, 
and the balance equations for $i+j\geq m$, remain the same.

\section{Numerical experiments}

The first question we wish to address is how the two-phase 
Foreground-Back\-ground queuing policy without speed modulation 
(to be referred to as FB-Ph2) performs in comparison with other service 
disciplines. In particular, we shall compare it with the First-Come-First-Served 
policy (FCFS, also known as First-In-First-Out, FIFO), and with the 
Least-Attained-Service policy (LAS; also called Shortest-Elapsed-Time, SET). 
The reason for choosing those two is that FCFS is perhaps the commonest 
policy in use, while LAS is known to have a certain optimality property: it 
minimizes the steady-state average number of jobs, $L$, when the service 
time distribution has a decreasing failure rate and the policy is `blind' (i.e. 
the exact service times are not known in advance), e.g., see \cite{nuy}.

We use the expressions derived for the special case $K=1$ in order 
to compute $L=L_1+L_2$ under the FB-Ph2 policy. In the case of FCFS, 
the value of $L$ is given by the Pollaczek-Khinchin formula:
\begin{equation} \label{pk}
L = \rho + \frac{\lambda^2M_2}{2(1-\rho)}\;,
\end{equation}
where $\rho$ is the offered load and $M_2$ 
is the second moment of the service time distribution. For the 
two-phase Coxian distribution without speed modulation we have 
$\rho =\rho_1 + \rho_2q$, and
\begin{equation} \label{m2}
M_2 = \frac{2}{\mu_1^2}+\frac{2q}{\mu_1\mu_2}+\frac{2q}{\mu_2^2}\;.
\end{equation}

The average number of jobs under the LAS policy is given by Schrage's 
formula:
\begin{equation} \label{Lset}
L = \int_0^\infty f(x) \left [ \frac{x}{1-\rho(x)}+\frac{\lambda  M_2(x)}{2(1-\rho(x))^2} 
\right ]dx\;,
\end{equation}
where $f(x)$ is the density of the service time distribution;
\[
\rho(x) = \lambda \int_0^x [1-F(t)]dt\;,
\]
is the truncated offered load, with $F(t)$ the cumulative distribution 
function;
\[
M_2(x) = 2\int_0^x t[1-F(t)]dt\;,
\]
is the truncated second moment of the service time. 

The density of the two-stage Coxian distribution is 
\[
f(x) =\mu_1 e^{-mu_1x} + \frac{\mu_1q}{\mu_1-\mu_2} 
[\mu_2 e^{-mu_2x} - \mu_1 e^{-mu_1x}]\;.
\]
The cumulative distribution and truncated moments are easily derived.

\begin{figure}[!ht]
	\begin{center}
	% GNUPLOT: LaTeX picture
	\setlength{\unitlength}{0.2pt}
	\ifx\plotpoint\undefined\newsavebox{\plotpoint}\fi
	\sbox{\plotpoint}{\rule[-0.200pt]{0.400pt}{0.400pt}}%
	\begin{picture}(1500,900)(0,0)
	\sbox{\plotpoint}{\rule[-0.200pt]{0.400pt}{0.400pt}}%
	\put(131.0,131.0){\rule[-0.200pt]{4.818pt}{0.400pt}}
	\put(111,131){\makebox(0,0)[r]{$0$}}
	\put(1419.0,131.0){\rule[-0.200pt]{4.818pt}{0.400pt}}
	\put(131.0,212.0){\rule[-0.200pt]{4.818pt}{0.400pt}}
	\put(111,212){\makebox(0,0)[r]{$5$}}
	\put(1419.0,212.0){\rule[-0.200pt]{4.818pt}{0.400pt}}
	\put(131.0,293.0){\rule[-0.200pt]{4.818pt}{0.400pt}}
	\put(111,293){\makebox(0,0)[r]{$10$}}
	\put(1419.0,293.0){\rule[-0.200pt]{4.818pt}{0.400pt}}
	\put(131.0,373.0){\rule[-0.200pt]{4.818pt}{0.400pt}}
	\put(111,373){\makebox(0,0)[r]{$15$}}
	\put(1419.0,373.0){\rule[-0.200pt]{4.818pt}{0.400pt}}
	\put(131.0,454.0){\rule[-0.200pt]{4.818pt}{0.400pt}}
	\put(111,454){\makebox(0,0)[r]{$20$}}
	\put(1419.0,454.0){\rule[-0.200pt]{4.818pt}{0.400pt}}
	\put(131.0,535.0){\rule[-0.200pt]{4.818pt}{0.400pt}}
	\put(111,535){\makebox(0,0)[r]{$25$}}
	\put(1419.0,535.0){\rule[-0.200pt]{4.818pt}{0.400pt}}
	\put(131.0,616.0){\rule[-0.200pt]{4.818pt}{0.400pt}}
	\put(111,616){\makebox(0,0)[r]{$30$}}
	\put(1419.0,616.0){\rule[-0.200pt]{4.818pt}{0.400pt}}
	\put(131.0,696.0){\rule[-0.200pt]{4.818pt}{0.400pt}}
	\put(111,696){\makebox(0,0)[r]{$35$}}
	\put(1419.0,696.0){\rule[-0.200pt]{4.818pt}{0.400pt}}
	\put(131.0,777.0){\rule[-0.200pt]{4.818pt}{0.400pt}}
	\put(111,777){\makebox(0,0)[r]{$40$}}
	\put(1419.0,777.0){\rule[-0.200pt]{4.818pt}{0.400pt}}
	\put(131.0,858.0){\rule[-0.200pt]{4.818pt}{0.400pt}}
	\put(111,858){\makebox(0,0)[r]{$45$}}
	\put(1419.0,858.0){\rule[-0.200pt]{4.818pt}{0.400pt}}
	\put(131.0,131.0){\rule[-0.200pt]{0.400pt}{4.818pt}}
	\put(131,90){\makebox(0,0){$2$}}
	\put(131.0,838.0){\rule[-0.200pt]{0.400pt}{4.818pt}}
	\put(349.0,131.0){\rule[-0.200pt]{0.400pt}{4.818pt}}
	\put(349,90){\makebox(0,0){$2.2$}}
	\put(349.0,838.0){\rule[-0.200pt]{0.400pt}{4.818pt}}
	\put(567.0,131.0){\rule[-0.200pt]{0.400pt}{4.818pt}}
	\put(567,90){\makebox(0,0){$2.4$}}
	\put(567.0,838.0){\rule[-0.200pt]{0.400pt}{4.818pt}}
	\put(785.0,131.0){\rule[-0.200pt]{0.400pt}{4.818pt}}
	\put(785,90){\makebox(0,0){$2.6$}}
	\put(785.0,838.0){\rule[-0.200pt]{0.400pt}{4.818pt}}
	\put(1003.0,131.0){\rule[-0.200pt]{0.400pt}{4.818pt}}
	\put(1003,90){\makebox(0,0){$2.8$}}
	\put(1003.0,838.0){\rule[-0.200pt]{0.400pt}{4.818pt}}
	\put(1221.0,131.0){\rule[-0.200pt]{0.400pt}{4.818pt}}
	\put(1221,90){\makebox(0,0){$3$}}
	\put(1221.0,838.0){\rule[-0.200pt]{0.400pt}{4.818pt}}
	\put(1439.0,131.0){\rule[-0.200pt]{0.400pt}{4.818pt}}
	\put(1439,90){\makebox(0,0){$3.2$}}
	\put(1439.0,838.0){\rule[-0.200pt]{0.400pt}{4.818pt}}
	%\put(131.0,131.0){\rule[-0.200pt]{0.400pt}{175.134pt}}
    %\put(131.0,131.0){\rule[-0.200pt]{315.097pt}{0.400pt}}
    %\put(1439.0,131.0){\rule[-0.200pt]{0.400pt}{175.134pt}}
    %\put(131.0,858.0){\rule[-0.200pt]{315.097pt}{0.400pt}}
	\put(131.0,131.0){\rule[-0.200pt]{0.400pt}{145pt}}
	\put(131.0,131.0){\rule[-0.200pt]{261pt}{0.400pt}}
	\put(1439.0,131.0){\rule[-0.200pt]{0.400pt}{145pt}}
	\put(131.0,858.0){\rule[-0.200pt]{261pt}{0.400pt}}
	\put(36,494){\makebox(0,0){$L$}}
	\put(785,29){\makebox(0,0){$\lambda$}}
	\put(511,817){\makebox(0,0)[r]{FCFS}}
	\put(531.0,817.0){\rule[-0.200pt]{24.090pt}{0.400pt}}
	\put(240,172){\usebox{\plotpoint}}
	\multiput(240.00,172.59)(9.803,0.482){9}{\rule{7.367pt}{0.116pt}}
	\multiput(240.00,171.17)(93.710,6.000){2}{\rule{3.683pt}{0.400pt}}
	\multiput(349.00,178.59)(7.163,0.488){13}{\rule{5.550pt}{0.117pt}}
	\multiput(349.00,177.17)(97.481,8.000){2}{\rule{2.775pt}{0.400pt}}
	\multiput(458.00,186.58)(5.655,0.491){17}{\rule{4.460pt}{0.118pt}}
	\multiput(458.00,185.17)(99.743,10.000){2}{\rule{2.230pt}{0.400pt}}
	\multiput(567.00,196.58)(4.675,0.492){21}{\rule{3.733pt}{0.119pt}}
	\multiput(567.00,195.17)(101.251,12.000){2}{\rule{1.867pt}{0.400pt}}
	\multiput(676.00,208.58)(3.712,0.494){27}{\rule{3.007pt}{0.119pt}}
	\multiput(676.00,207.17)(102.760,15.000){2}{\rule{1.503pt}{0.400pt}}
	\multiput(785.00,223.58)(2.765,0.496){37}{\rule{2.280pt}{0.119pt}}
	\multiput(785.00,222.17)(104.268,20.000){2}{\rule{1.140pt}{0.400pt}}
	\multiput(894.00,243.58)(1.964,0.497){53}{\rule{1.657pt}{0.120pt}}
	\multiput(894.00,242.17)(105.561,28.000){2}{\rule{0.829pt}{0.400pt}}
	\multiput(1003.00,271.58)(1.336,0.498){79}{\rule{1.163pt}{0.120pt}}
	\multiput(1003.00,270.17)(106.585,41.000){2}{\rule{0.582pt}{0.400pt}}
	\multiput(1112.00,312.58)(0.827,0.499){129}{\rule{0.761pt}{0.120pt}}
	\multiput(1112.00,311.17)(107.421,66.000){2}{\rule{0.380pt}{0.400pt}}
	\multiput(1221.58,378.00)(0.499,0.564){215}{\rule{0.120pt}{0.551pt}}
	\multiput(1220.17,378.00)(109.000,121.856){2}{\rule{0.400pt}{0.276pt}}
	\multiput(1330.58,501.00)(0.499,1.411){215}{\rule{0.120pt}{1.227pt}}
	\multiput(1329.17,501.00)(109.000,304.454){2}{\rule{0.400pt}{0.613pt}}
	\put(240,172){\makebox(0,0){$+$}}
	\put(349,178){\makebox(0,0){$+$}}
	\put(458,186){\makebox(0,0){$+$}}
	\put(567,196){\makebox(0,0){$+$}}
	\put(676,208){\makebox(0,0){$+$}}
	\put(785,223){\makebox(0,0){$+$}}
	\put(894,243){\makebox(0,0){$+$}}
	\put(1003,271){\makebox(0,0){$+$}}
	\put(1112,312){\makebox(0,0){$+$}}
	\put(1221,378){\makebox(0,0){$+$}}
	\put(1330,501){\makebox(0,0){$+$}}
	\put(1439,808){\makebox(0,0){$+$}}
	\put(581,817){\makebox(0,0){$+$}}
	\put(511,776){\makebox(0,0)[r]{LAS}}
	\multiput(531,776)(20.756,0.000){5}{\usebox{\plotpoint}}
	\put(631,776){\usebox{\plotpoint}}
	\put(240,155){\usebox{\plotpoint}}
	\multiput(240,155)(20.748,0.571){6}{\usebox{\plotpoint}}
	\multiput(349,158)(20.742,0.761){5}{\usebox{\plotpoint}}
	\multiput(458,162)(20.742,0.761){5}{\usebox{\plotpoint}}
	\multiput(567,166)(20.734,0.951){6}{\usebox{\plotpoint}}
	\multiput(676,171)(20.724,1.141){5}{\usebox{\plotpoint}}
	\multiput(785,177)(20.700,1.519){5}{\usebox{\plotpoint}}
	\multiput(894,185)(20.669,1.896){5}{\usebox{\plotpoint}}
	\multiput(1003,195)(20.562,2.830){6}{\usebox{\plotpoint}}
	\multiput(1112,210)(20.308,4.285){5}{\usebox{\plotpoint}}
	\multiput(1221,233)(19.485,7.150){6}{\usebox{\plotpoint}}
	\multiput(1330,273)(15.505,13.798){7}{\usebox{\plotpoint}}
	\put(1439,370){\usebox{\plotpoint}}
	\put(240,155){\makebox(0,0){$\times$}}
	\put(349,158){\makebox(0,0){$\times$}}
	\put(458,162){\makebox(0,0){$\times$}}
	\put(567,166){\makebox(0,0){$\times$}}
	\put(676,171){\makebox(0,0){$\times$}}
	\put(785,177){\makebox(0,0){$\times$}}
	\put(894,185){\makebox(0,0){$\times$}}
	\put(1003,195){\makebox(0,0){$\times$}}
	\put(1112,210){\makebox(0,0){$\times$}}
	\put(1221,233){\makebox(0,0){$\times$}}
	\put(1330,273){\makebox(0,0){$\times$}}
	\put(1439,370){\makebox(0,0){$\times$}}
	\put(581,776){\makebox(0,0){$\times$}}
	\sbox{\plotpoint}{\rule[-0.400pt]{0.800pt}{0.800pt}}%
	\sbox{\plotpoint}{\rule[-0.200pt]{0.400pt}{0.400pt}}%
	\put(511,735){\makebox(0,0)[r]{FB-ph2}}
	\sbox{\plotpoint}{\rule[-0.400pt]{0.800pt}{0.800pt}}%
	\put(531.0,735.0){\rule[-0.400pt]{24.090pt}{0.800pt}}
	\put(240,154){\usebox{\plotpoint}}
	\put(240,153.84){\rule{26.258pt}{0.800pt}}
	\multiput(240.00,152.34)(54.500,3.000){2}{\rule{13.129pt}{0.800pt}}
	\put(349,156.84){\rule{26.258pt}{0.800pt}}
	\multiput(349.00,155.34)(54.500,3.000){2}{\rule{13.129pt}{0.800pt}}
	\put(458,160.34){\rule{22.000pt}{0.800pt}}
	\multiput(458.00,158.34)(63.338,4.000){2}{\rule{11.000pt}{0.800pt}}
	\put(567,164.34){\rule{22.000pt}{0.800pt}}
	\multiput(567.00,162.34)(63.338,4.000){2}{\rule{11.000pt}{0.800pt}}
	\multiput(676.00,169.39)(11.960,0.536){5}{\rule{14.733pt}{0.129pt}}
	\multiput(676.00,166.34)(78.420,6.000){2}{\rule{7.367pt}{0.800pt}}
	\multiput(785.00,175.40)(9.410,0.526){7}{\rule{12.657pt}{0.127pt}}
	\multiput(785.00,172.34)(82.729,7.000){2}{\rule{6.329pt}{0.800pt}}
	\multiput(894.00,182.40)(5.998,0.514){13}{\rule{8.920pt}{0.124pt}}
	\multiput(894.00,179.34)(90.486,10.000){2}{\rule{4.460pt}{0.800pt}}
	\multiput(1003.00,192.41)(4.463,0.509){19}{\rule{6.908pt}{0.123pt}}
	\multiput(1003.00,189.34)(94.663,13.000){2}{\rule{3.454pt}{0.800pt}}
	\multiput(1112.00,205.41)(2.548,0.505){37}{\rule{4.164pt}{0.122pt}}
	\multiput(1112.00,202.34)(100.358,22.000){2}{\rule{2.082pt}{0.800pt}}
	\multiput(1221.00,227.41)(1.451,0.503){69}{\rule{2.495pt}{0.121pt}}
	\multiput(1221.00,224.34)(103.822,38.000){2}{\rule{1.247pt}{0.800pt}}
	\multiput(1330.00,265.41)(0.580,0.501){181}{\rule{1.128pt}{0.121pt}}
	\multiput(1330.00,262.34)(106.659,94.000){2}{\rule{0.564pt}{0.800pt}}
	\put(240,154){\makebox(0,0){$\ast$}}
	\put(349,157){\makebox(0,0){$\ast$}}
	\put(458,160){\makebox(0,0){$\ast$}}
	\put(567,164){\makebox(0,0){$\ast$}}
	\put(676,168){\makebox(0,0){$\ast$}}
	\put(785,174){\makebox(0,0){$\ast$}}
	\put(894,181){\makebox(0,0){$\ast$}}
	\put(1003,191){\makebox(0,0){$\ast$}}
	\put(1112,204){\makebox(0,0){$\ast$}}
	\put(1221,226){\makebox(0,0){$\ast$}}
	\put(1330,264){\makebox(0,0){$\ast$}}
	\put(1439,358){\makebox(0,0){$\ast$}}
	\put(581,735){\makebox(0,0){$\ast$}}
	\sbox{\plotpoint}{\rule[-0.200pt]{0.400pt}{0.400pt}}%
	\end{picture}
	\caption{ Comparison between three policies} \label{f1} 
	$\mu_1=5$, $\mu_2=1$, $q=0.1$
	\end{center}
\end{figure}
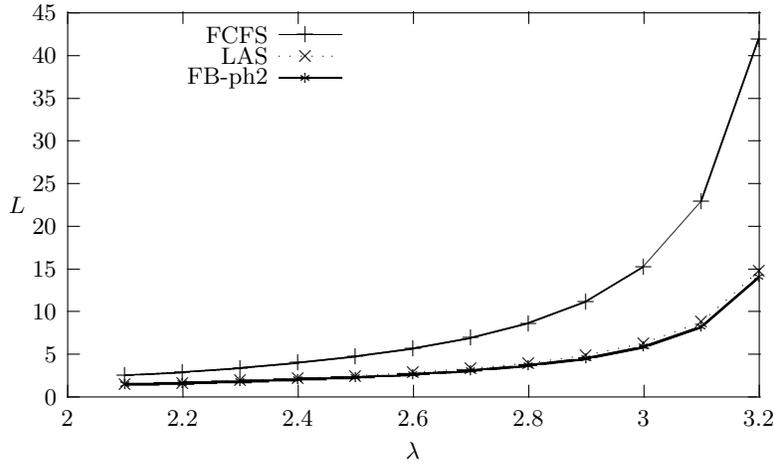

We have taken an example where $\mu_1=5$, $\mu_2=1$ and $q=0.1$. That is, 
the second phase does not occur frequently but when it does, it is 
five times longer, on the average, than the first. In Figure \ref{f1}, 
the values of $L$ under the three service policies are plotted against 
the arrival rate, $\lambda$. The offered load $\rho$ varies from about 
60\% to about 95\%.

Unsurprisingly, the FCFS policy has the worst performance, with the fastest 
increase in the average queue size. On the other hand, the other two plots 
appear to be counter-intuitive. One would expect to find that the optimal 
policy, LAS, has the best performance, while FB-ph2, which also favours 
short jobs, performs a little worse. Instead, we observe that FB-ph2 is slightly 
better than LAS. This `contradiction' of the optimality of LAS is 
explained by the fact that although the service time distribution has a 
decreasing failure rate, the policy is not quite 'blind'. It is not aware of 
exact service times but it does know that after the short first phase a job 
will leave the system with probability $1-q$. It seems that running first 
phases to completion is better than allowing them to be interrupted by new 
arrivals.

\begin{figure}[!ht]
	\begin{center}
		% GNUPLOT: LaTeX picture
		\setlength{\unitlength}{0.2pt}
		\ifx\plotpoint\undefined\newsavebox{\plotpoint}\fi
		\sbox{\plotpoint}{\rule[-0.200pt]{0.400pt}{0.400pt}}%
\begin{picture}(1500,900)(0,0)
	\sbox{\plotpoint}{\rule[-0.200pt]{0.400pt}{0.400pt}}%
	\put(131.0,131.0){\rule[-0.200pt]{4.818pt}{0.400pt}}
	\put(111,131){\makebox(0,0)[r]{$0$}}
	\put(1419.0,131.0){\rule[-0.200pt]{4.818pt}{0.400pt}}
	\put(131.0,235.0){\rule[-0.200pt]{4.818pt}{0.400pt}}
	\put(111,235){\makebox(0,0)[r]{$10$}}
	\put(1419.0,235.0){\rule[-0.200pt]{4.818pt}{0.400pt}}
	\put(131.0,339.0){\rule[-0.200pt]{4.818pt}{0.400pt}}
	\put(111,339){\makebox(0,0)[r]{$20$}}
	\put(1419.0,339.0){\rule[-0.200pt]{4.818pt}{0.400pt}}
	\put(131.0,443.0){\rule[-0.200pt]{4.818pt}{0.400pt}}
	\put(111,443){\makebox(0,0)[r]{$30$}}
	\put(1419.0,443.0){\rule[-0.200pt]{4.818pt}{0.400pt}}
	\put(131.0,546.0){\rule[-0.200pt]{4.818pt}{0.400pt}}
	\put(111,546){\makebox(0,0)[r]{$40$}}
	\put(1419.0,546.0){\rule[-0.200pt]{4.818pt}{0.400pt}}
	\put(131.0,650.0){\rule[-0.200pt]{4.818pt}{0.400pt}}
	\put(111,650){\makebox(0,0)[r]{$50$}}
	\put(1419.0,650.0){\rule[-0.200pt]{4.818pt}{0.400pt}}
	\put(131.0,754.0){\rule[-0.200pt]{4.818pt}{0.400pt}}
	\put(111,754){\makebox(0,0)[r]{$60$}}
	\put(1419.0,754.0){\rule[-0.200pt]{4.818pt}{0.400pt}}
	\put(131.0,858.0){\rule[-0.200pt]{4.818pt}{0.400pt}}
	\put(111,858){\makebox(0,0)[r]{$70$}}
	\put(1419.0,858.0){\rule[-0.200pt]{4.818pt}{0.400pt}}
	\put(131.0,131.0){\rule[-0.200pt]{0.400pt}{4.818pt}}
	\put(131,90){\makebox(0,0){$2$}}
	\put(131.0,838.0){\rule[-0.200pt]{0.400pt}{4.818pt}}
	\put(349.0,131.0){\rule[-0.200pt]{0.400pt}{4.818pt}}
	\put(349,90){\makebox(0,0){$2.2$}}
	\put(349.0,838.0){\rule[-0.200pt]{0.400pt}{4.818pt}}
	\put(567.0,131.0){\rule[-0.200pt]{0.400pt}{4.818pt}}
	\put(567,90){\makebox(0,0){$2.4$}}
	\put(567.0,838.0){\rule[-0.200pt]{0.400pt}{4.818pt}}
	\put(785.0,131.0){\rule[-0.200pt]{0.400pt}{4.818pt}}
	\put(785,90){\makebox(0,0){$2.6$}}
	\put(785.0,838.0){\rule[-0.200pt]{0.400pt}{4.818pt}}
	\put(1003.0,131.0){\rule[-0.200pt]{0.400pt}{4.818pt}}
	\put(1003,90){\makebox(0,0){$2.8$}}
	\put(1003.0,838.0){\rule[-0.200pt]{0.400pt}{4.818pt}}
	\put(1221.0,131.0){\rule[-0.200pt]{0.400pt}{4.818pt}}
	\put(1221,90){\makebox(0,0){$3$}}
	\put(1221.0,838.0){\rule[-0.200pt]{0.400pt}{4.818pt}}
	\put(1439.0,131.0){\rule[-0.200pt]{0.400pt}{4.818pt}}
	\put(1439,90){\makebox(0,0){$3.2$}}
	\put(1439.0,838.0){\rule[-0.200pt]{0.400pt}{4.818pt}}
\put(131.0,131.0){\rule[-0.200pt]{0.400pt}{145pt}}
\put(131.0,131.0){\rule[-0.200pt]{261pt}{0.400pt}}
\put(1439.0,131.0){\rule[-0.200pt]{0.400pt}{145pt}}
\put(131.0,858.0){\rule[-0.200pt]{261pt}{0.400pt}}
	\put(36,494){\makebox(0,0){$L$}}
	\put(785,29){\makebox(0,0){$\lambda$}}
	\put(531,817){\makebox(0,0)[r]{FCFS}}
	\put(551.0,817.0){\rule[-0.200pt]{24.090pt}{0.400pt}}
	\put(240,170){\usebox{\plotpoint}}
	\multiput(240.00,170.59)(9.803,0.482){9}{\rule{7.367pt}{0.116pt}}
	\multiput(240.00,169.17)(93.710,6.000){2}{\rule{3.683pt}{0.400pt}}
	\multiput(349.00,176.59)(7.163,0.488){13}{\rule{5.550pt}{0.117pt}}
	\multiput(349.00,175.17)(97.481,8.000){2}{\rule{2.775pt}{0.400pt}}
	\multiput(458.00,184.58)(5.655,0.491){17}{\rule{4.460pt}{0.118pt}}
	\multiput(458.00,183.17)(99.743,10.000){2}{\rule{2.230pt}{0.400pt}}
	\multiput(567.00,194.58)(4.675,0.492){21}{\rule{3.733pt}{0.119pt}}
	\multiput(567.00,193.17)(101.251,12.000){2}{\rule{1.867pt}{0.400pt}}
	\multiput(676.00,206.58)(3.474,0.494){29}{\rule{2.825pt}{0.119pt}}
	\multiput(676.00,205.17)(103.137,16.000){2}{\rule{1.413pt}{0.400pt}}
	\multiput(785.00,222.58)(2.631,0.496){39}{\rule{2.176pt}{0.119pt}}
	\multiput(785.00,221.17)(104.483,21.000){2}{\rule{1.088pt}{0.400pt}}
	\multiput(894.00,243.58)(1.895,0.497){55}{\rule{1.603pt}{0.120pt}}
	\multiput(894.00,242.17)(105.672,29.000){2}{\rule{0.802pt}{0.400pt}}
	\multiput(1003.00,272.58)(1.273,0.498){83}{\rule{1.114pt}{0.120pt}}
	\multiput(1003.00,271.17)(106.688,43.000){2}{\rule{0.557pt}{0.400pt}}
	\multiput(1112.00,315.58)(0.802,0.499){133}{\rule{0.741pt}{0.120pt}}
	\multiput(1112.00,314.17)(107.462,68.000){2}{\rule{0.371pt}{0.400pt}}
	\multiput(1221.58,383.00)(0.499,0.587){215}{\rule{0.120pt}{0.570pt}}
	\multiput(1220.17,383.00)(109.000,126.818){2}{\rule{0.400pt}{0.285pt}}
	\multiput(1330.58,511.00)(0.499,1.475){215}{\rule{0.120pt}{1.278pt}}
	\multiput(1329.17,511.00)(109.000,318.347){2}{\rule{0.400pt}{0.639pt}}
	\put(240,170){\makebox(0,0){$+$}}
	\put(349,176){\makebox(0,0){$+$}}
	\put(458,184){\makebox(0,0){$+$}}
	\put(567,194){\makebox(0,0){$+$}}
	\put(676,206){\makebox(0,0){$+$}}
	\put(785,222){\makebox(0,0){$+$}}
	\put(894,243){\makebox(0,0){$+$}}
	\put(1003,272){\makebox(0,0){$+$}}
	\put(1112,315){\makebox(0,0){$+$}}
	\put(1221,383){\makebox(0,0){$+$}}
	\put(1330,511){\makebox(0,0){$+$}}
	\put(1439,832){\makebox(0,0){$+$}}
	\put(601,817){\makebox(0,0){$+$}}
	\put(531,776){\makebox(0,0)[r]{LAS}}
	\multiput(551,776)(20.756,0.000){5}{\usebox{\plotpoint}}
	\put(651,776){\usebox{\plotpoint}}
	\put(240,146){\usebox{\plotpoint}}
	\multiput(240,146)(20.755,0.190){6}{\usebox{\plotpoint}}
	\multiput(349,147)(20.752,0.381){5}{\usebox{\plotpoint}}
	\multiput(458,149)(20.748,0.571){5}{\usebox{\plotpoint}}
	\multiput(567,152)(20.752,0.381){6}{\usebox{\plotpoint}}
	\multiput(676,154)(20.742,0.761){5}{\usebox{\plotpoint}}
	\multiput(785,158)(20.742,0.761){5}{\usebox{\plotpoint}}
	\multiput(894,162)(20.724,1.141){5}{\usebox{\plotpoint}}
	\multiput(1003,168)(20.713,1.330){6}{\usebox{\plotpoint}}
	\multiput(1112,175)(20.631,2.271){5}{\usebox{\plotpoint}}
	\multiput(1221,187)(20.381,3.927){5}{\usebox{\plotpoint}}
	\multiput(1330,208)(18.799,8.796){6}{\usebox{\plotpoint}}
	\put(1439,259){\usebox{\plotpoint}}
	\put(240,146){\makebox(0,0){$\times$}}
	\put(349,147){\makebox(0,0){$\times$}}
	\put(458,149){\makebox(0,0){$\times$}}
	\put(567,152){\makebox(0,0){$\times$}}
	\put(676,154){\makebox(0,0){$\times$}}
	\put(785,158){\makebox(0,0){$\times$}}
	\put(894,162){\makebox(0,0){$\times$}}
	\put(1003,168){\makebox(0,0){$\times$}}
	\put(1112,175){\makebox(0,0){$\times$}}
	\put(1221,187){\makebox(0,0){$\times$}}
	\put(1330,208){\makebox(0,0){$\times$}}
	\put(1439,259){\makebox(0,0){$\times$}}
	\put(601,776){\makebox(0,0){$\times$}}
	\sbox{\plotpoint}{\rule[-0.400pt]{0.800pt}{0.800pt}}%
	\sbox{\plotpoint}{\rule[-0.200pt]{0.400pt}{0.400pt}}%
	\put(531,735){\makebox(0,0)[r]{FB-ph2}}
	\sbox{\plotpoint}{\rule[-0.400pt]{0.800pt}{0.800pt}}%
	\put(551.0,735.0){\rule[-0.400pt]{24.090pt}{0.800pt}}
	\put(240,145){\usebox{\plotpoint}}
	\put(240,144.34){\rule{26.258pt}{0.800pt}}
	\multiput(240.00,143.34)(54.500,2.000){2}{\rule{13.129pt}{0.800pt}}
	\put(349,146.34){\rule{26.258pt}{0.800pt}}
	\multiput(349.00,145.34)(54.500,2.000){2}{\rule{13.129pt}{0.800pt}}
	\put(458,148.34){\rule{26.258pt}{0.800pt}}
	\multiput(458.00,147.34)(54.500,2.000){2}{\rule{13.129pt}{0.800pt}}
	\put(567,150.34){\rule{26.258pt}{0.800pt}}
	\multiput(567.00,149.34)(54.500,2.000){2}{\rule{13.129pt}{0.800pt}}
	\put(676,153.34){\rule{22.000pt}{0.800pt}}
	\multiput(676.00,151.34)(63.338,4.000){2}{\rule{11.000pt}{0.800pt}}
	\put(785,157.34){\rule{22.000pt}{0.800pt}}
	\multiput(785.00,155.34)(63.338,4.000){2}{\rule{11.000pt}{0.800pt}}
	\multiput(894.00,162.38)(17.887,0.560){3}{\rule{17.640pt}{0.135pt}}
	\multiput(894.00,159.34)(72.387,5.000){2}{\rule{8.820pt}{0.800pt}}
	\multiput(1003.00,167.40)(9.410,0.526){7}{\rule{12.657pt}{0.127pt}}
	\multiput(1003.00,164.34)(82.729,7.000){2}{\rule{6.329pt}{0.800pt}}
	\multiput(1112.00,174.41)(4.875,0.511){17}{\rule{7.467pt}{0.123pt}}
	\multiput(1112.00,171.34)(93.503,12.000){2}{\rule{3.733pt}{0.800pt}}
	\multiput(1221.00,186.41)(2.815,0.505){33}{\rule{4.560pt}{0.122pt}}
	\multiput(1221.00,183.34)(99.535,20.000){2}{\rule{2.280pt}{0.800pt}}
	\multiput(1330.00,206.41)(1.097,0.502){93}{\rule{1.944pt}{0.121pt}}
	\multiput(1330.00,203.34)(104.965,50.000){2}{\rule{0.972pt}{0.800pt}}
	\put(240,145){\makebox(0,0){$\ast$}}
	\put(349,147){\makebox(0,0){$\ast$}}
	\put(458,149){\makebox(0,0){$\ast$}}
	\put(567,151){\makebox(0,0){$\ast$}}
	\put(676,153){\makebox(0,0){$\ast$}}
	\put(785,157){\makebox(0,0){$\ast$}}
	\put(894,161){\makebox(0,0){$\ast$}}
	\put(1003,166){\makebox(0,0){$\ast$}}
	\put(1112,173){\makebox(0,0){$\ast$}}
	\put(1221,185){\makebox(0,0){$\ast$}}
	\put(1330,205){\makebox(0,0){$\ast$}}
	\put(1439,255){\makebox(0,0){$\ast$}}
	\put(601,735){\makebox(0,0){$\ast$}}
	\sbox{\plotpoint}{\rule[-0.200pt]{0.400pt}{0.400pt}}%
%	\put(131.0,131.0){\rule[-0.200pt]{0.400pt}{175.134pt}}
%	\put(131.0,131.0){\rule[-0.200pt]{315.097pt}{0.400pt}}
%	\put(1439.0,131.0){\rule[-0.200pt]{0.400pt}{175.134pt}}
%	\put(131.0,858.0){\rule[-0.200pt]{315.097pt}{0.400pt}}
\end{picture}
			\caption{ Comparison between three policies} \label{f3a} 
		$\mu_1=5$, $\mu_2=0.5$, $q=0.05$
	\end{center}
\end{figure}
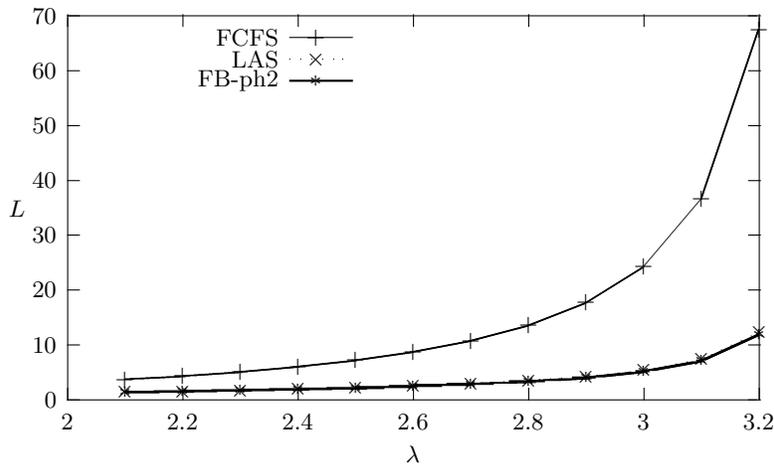

In order to examine this phenomenon further, we have replicated the 
experiment twice, keeping $\mu_1$ fixed but varying $\mu_2$ and $q$ in 
proportion. That is, the average required service time remains the same but 
its coefficient of variation changes. First, $\mu_2$ and $q$ are both halved: 
$\mu_2=0.5$, $q=0.05$. This increases the coefficient of variation. The 
resulting comparison is shown in Figure \ref{f3a}

Predictably, the performance of the FCFS policy is now considerably 
worse. Less predictably, both LAS and FB-ph2 perform better than 
before, and while FB-ph2 is still slightly better than LAS, the difference 
between them has decreased. These observations can be explaned by
noting that the coupling between phase 1 and phase 2 becomes less 
important when second phases are rare. If $\mu_2$ and $q$ are small, 
both LAS and FB-ph2 behave like a two-class preemptive priority system 
with independent Poisson arrival streams, with rates $\lambda$ and 
$\lambda q$, respectively. Indeed, evaluating the value of $L$ in the 
latter system using the known solution, produces results thst are very 
close to the ones in the figure. Thus, weaker coupling explains both the 
improved performance and the reduced distance between the two plots. 

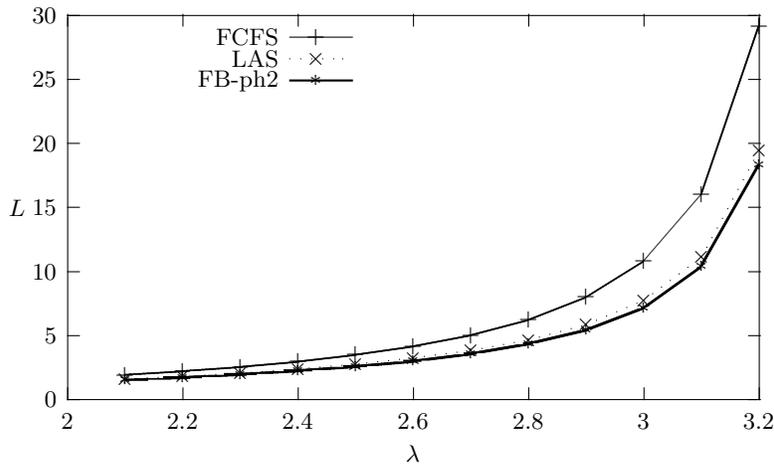
\begin{figure}[!ht]
	\begin{center}
		% GNUPLOT: LaTeX picture
		\setlength{\unitlength}{0.2pt}
		\ifx\plotpoint\undefined\newsavebox{\plotpoint}\fi
		\sbox{\plotpoint}{\rule[-0.200pt]{0.400pt}{0.400pt}}%
\begin{picture}(1500,900)(0,0)
	\sbox{\plotpoint}{\rule[-0.200pt]{0.400pt}{0.400pt}}%
	\put(131.0,131.0){\rule[-0.200pt]{4.818pt}{0.400pt}}
	\put(111,131){\makebox(0,0)[r]{$0$}}
	\put(1419.0,131.0){\rule[-0.200pt]{4.818pt}{0.400pt}}
	\put(131.0,252.0){\rule[-0.200pt]{4.818pt}{0.400pt}}
	\put(111,252){\makebox(0,0)[r]{$5$}}
	\put(1419.0,252.0){\rule[-0.200pt]{4.818pt}{0.400pt}}
	\put(131.0,373.0){\rule[-0.200pt]{4.818pt}{0.400pt}}
	\put(111,373){\makebox(0,0)[r]{$10$}}
	\put(1419.0,373.0){\rule[-0.200pt]{4.818pt}{0.400pt}}
	\put(131.0,495.0){\rule[-0.200pt]{4.818pt}{0.400pt}}
	\put(111,495){\makebox(0,0)[r]{$15$}}
	\put(1419.0,495.0){\rule[-0.200pt]{4.818pt}{0.400pt}}
	\put(131.0,616.0){\rule[-0.200pt]{4.818pt}{0.400pt}}
	\put(111,616){\makebox(0,0)[r]{$20$}}
	\put(1419.0,616.0){\rule[-0.200pt]{4.818pt}{0.400pt}}
	\put(131.0,737.0){\rule[-0.200pt]{4.818pt}{0.400pt}}
	\put(111,737){\makebox(0,0)[r]{$25$}}
	\put(1419.0,737.0){\rule[-0.200pt]{4.818pt}{0.400pt}}
	\put(131.0,858.0){\rule[-0.200pt]{4.818pt}{0.400pt}}
	\put(111,858){\makebox(0,0)[r]{$30$}}
	\put(1419.0,858.0){\rule[-0.200pt]{4.818pt}{0.400pt}}
	\put(131.0,131.0){\rule[-0.200pt]{0.400pt}{4.818pt}}
	\put(131,90){\makebox(0,0){$2$}}
	\put(131.0,838.0){\rule[-0.200pt]{0.400pt}{4.818pt}}
	\put(349.0,131.0){\rule[-0.200pt]{0.400pt}{4.818pt}}
	\put(349,90){\makebox(0,0){$2.2$}}
	\put(349.0,838.0){\rule[-0.200pt]{0.400pt}{4.818pt}}
	\put(567.0,131.0){\rule[-0.200pt]{0.400pt}{4.818pt}}
	\put(567,90){\makebox(0,0){$2.4$}}
	\put(567.0,838.0){\rule[-0.200pt]{0.400pt}{4.818pt}}
	\put(785.0,131.0){\rule[-0.200pt]{0.400pt}{4.818pt}}
	\put(785,90){\makebox(0,0){$2.6$}}
	\put(785.0,838.0){\rule[-0.200pt]{0.400pt}{4.818pt}}
	\put(1003.0,131.0){\rule[-0.200pt]{0.400pt}{4.818pt}}
	\put(1003,90){\makebox(0,0){$2.8$}}
	\put(1003.0,838.0){\rule[-0.200pt]{0.400pt}{4.818pt}}
	\put(1221.0,131.0){\rule[-0.200pt]{0.400pt}{4.818pt}}
	\put(1221,90){\makebox(0,0){$3$}}
	\put(1221.0,838.0){\rule[-0.200pt]{0.400pt}{4.818pt}}
	\put(1439.0,131.0){\rule[-0.200pt]{0.400pt}{4.818pt}}
	\put(1439,90){\makebox(0,0){$3.2$}}
	\put(1439.0,838.0){\rule[-0.200pt]{0.400pt}{4.818pt}}
\put(131.0,131.0){\rule[-0.200pt]{0.400pt}{145pt}}
\put(131.0,131.0){\rule[-0.200pt]{261pt}{0.400pt}}
\put(1439.0,131.0){\rule[-0.200pt]{0.400pt}{145pt}}
\put(131.0,858.0){\rule[-0.200pt]{261pt}{0.400pt}}
	\put(36,494){\makebox(0,0){$L$}}
	\put(785,29){\makebox(0,0){$\lambda$}}
	\put(531,817){\makebox(0,0)[r]{FCFS}}
	\put(551.0,817.0){\rule[-0.200pt]{24.090pt}{0.400pt}}
	\put(240,178){\usebox{\plotpoint}}
	\multiput(240.00,178.59)(8.273,0.485){11}{\rule{6.329pt}{0.117pt}}
	\multiput(240.00,177.17)(95.865,7.000){2}{\rule{3.164pt}{0.400pt}}
	\multiput(349.00,185.59)(7.163,0.488){13}{\rule{5.550pt}{0.117pt}}
	\multiput(349.00,184.17)(97.481,8.000){2}{\rule{2.775pt}{0.400pt}}
	\multiput(458.00,193.58)(5.655,0.491){17}{\rule{4.460pt}{0.118pt}}
	\multiput(458.00,192.17)(99.743,10.000){2}{\rule{2.230pt}{0.400pt}}
	\multiput(567.00,203.58)(4.303,0.493){23}{\rule{3.454pt}{0.119pt}}
	\multiput(567.00,202.17)(101.831,13.000){2}{\rule{1.727pt}{0.400pt}}
	\multiput(676.00,216.58)(3.474,0.494){29}{\rule{2.825pt}{0.119pt}}
	\multiput(676.00,215.17)(103.137,16.000){2}{\rule{1.413pt}{0.400pt}}
	\multiput(785.00,232.58)(2.631,0.496){39}{\rule{2.176pt}{0.119pt}}
	\multiput(785.00,231.17)(104.483,21.000){2}{\rule{1.088pt}{0.400pt}}
	\multiput(894.00,253.58)(1.895,0.497){55}{\rule{1.603pt}{0.120pt}}
	\multiput(894.00,252.17)(105.672,29.000){2}{\rule{0.802pt}{0.400pt}}
	\multiput(1003.00,282.58)(1.273,0.498){83}{\rule{1.114pt}{0.120pt}}
	\multiput(1003.00,281.17)(106.688,43.000){2}{\rule{0.557pt}{0.400pt}}
	\multiput(1112.00,325.58)(0.802,0.499){133}{\rule{0.741pt}{0.120pt}}
	\multiput(1112.00,324.17)(107.462,68.000){2}{\rule{0.371pt}{0.400pt}}
	\multiput(1221.58,393.00)(0.499,0.578){215}{\rule{0.120pt}{0.562pt}}
	\multiput(1220.17,393.00)(109.000,124.833){2}{\rule{0.400pt}{0.281pt}}
	\multiput(1330.58,519.00)(0.499,1.461){215}{\rule{0.120pt}{1.267pt}}
	\multiput(1329.17,519.00)(109.000,315.370){2}{\rule{0.400pt}{0.633pt}}
	\put(240,178){\makebox(0,0){$+$}}
	\put(349,185){\makebox(0,0){$+$}}
	\put(458,193){\makebox(0,0){$+$}}
	\put(567,203){\makebox(0,0){$+$}}
	\put(676,216){\makebox(0,0){$+$}}
	\put(785,232){\makebox(0,0){$+$}}
	\put(894,253){\makebox(0,0){$+$}}
	\put(1003,282){\makebox(0,0){$+$}}
	\put(1112,325){\makebox(0,0){$+$}}
	\put(1221,393){\makebox(0,0){$+$}}
	\put(1330,519){\makebox(0,0){$+$}}
	\put(1439,837){\makebox(0,0){$+$}}
	\put(601,817){\makebox(0,0){$+$}}
	\put(531,776){\makebox(0,0)[r]{LAS}}
	\multiput(551,776)(20.756,0.000){5}{\usebox{\plotpoint}}
	\put(651,776){\usebox{\plotpoint}}
	\put(240,170){\usebox{\plotpoint}}
	\multiput(240,170)(20.734,0.951){6}{\usebox{\plotpoint}}
	\multiput(349,175)(20.724,1.141){5}{\usebox{\plotpoint}}
	\multiput(458,181)(20.700,1.519){5}{\usebox{\plotpoint}}
	\multiput(567,189)(20.685,1.708){6}{\usebox{\plotpoint}}
	\multiput(676,198)(20.631,2.271){5}{\usebox{\plotpoint}}
	\multiput(785,210)(20.586,2.644){5}{\usebox{\plotpoint}}
	\multiput(894,224)(20.415,3.746){5}{\usebox{\plotpoint}}
	\multiput(1003,244)(20.058,5.336){6}{\usebox{\plotpoint}}
	\multiput(1112,273)(19.122,8.070){6}{\usebox{\plotpoint}}
	\multiput(1221,319)(16.586,12.478){6}{\usebox{\plotpoint}}
	\multiput(1330,401)(9.856,18.266){11}{\usebox{\plotpoint}}
	\put(1439,603){\usebox{\plotpoint}}
	\put(240,170){\makebox(0,0){$\times$}}
	\put(349,175){\makebox(0,0){$\times$}}
	\put(458,181){\makebox(0,0){$\times$}}
	\put(567,189){\makebox(0,0){$\times$}}
	\put(676,198){\makebox(0,0){$\times$}}
	\put(785,210){\makebox(0,0){$\times$}}
	\put(894,224){\makebox(0,0){$\times$}}
	\put(1003,244){\makebox(0,0){$\times$}}
	\put(1112,273){\makebox(0,0){$\times$}}
	\put(1221,319){\makebox(0,0){$\times$}}
	\put(1330,401){\makebox(0,0){$\times$}}
	\put(1439,603){\makebox(0,0){$\times$}}
	\put(601,776){\makebox(0,0){$\times$}}
	\sbox{\plotpoint}{\rule[-0.400pt]{0.800pt}{0.800pt}}%
	\sbox{\plotpoint}{\rule[-0.200pt]{0.400pt}{0.400pt}}%
	\put(531,735){\makebox(0,0)[r]{FB-ph2}}
	\sbox{\plotpoint}{\rule[-0.400pt]{0.800pt}{0.800pt}}%
	\put(551.0,735.0){\rule[-0.400pt]{24.090pt}{0.800pt}}
	\put(240,168){\usebox{\plotpoint}}
	\multiput(240.00,169.38)(17.887,0.560){3}{\rule{17.640pt}{0.135pt}}
	\multiput(240.00,166.34)(72.387,5.000){2}{\rule{8.820pt}{0.800pt}}
	\multiput(349.00,174.39)(11.960,0.536){5}{\rule{14.733pt}{0.129pt}}
	\multiput(349.00,171.34)(78.420,6.000){2}{\rule{7.367pt}{0.800pt}}
	\multiput(458.00,180.40)(9.410,0.526){7}{\rule{12.657pt}{0.127pt}}
	\multiput(458.00,177.34)(82.729,7.000){2}{\rule{6.329pt}{0.800pt}}
	\multiput(567.00,187.40)(7.865,0.520){9}{\rule{11.100pt}{0.125pt}}
	\multiput(567.00,184.34)(85.961,8.000){2}{\rule{5.550pt}{0.800pt}}
	\multiput(676.00,195.40)(5.998,0.514){13}{\rule{8.920pt}{0.124pt}}
	\multiput(676.00,192.34)(90.486,10.000){2}{\rule{4.460pt}{0.800pt}}
	\multiput(785.00,205.41)(4.116,0.509){21}{\rule{6.429pt}{0.123pt}}
	\multiput(785.00,202.34)(95.657,14.000){2}{\rule{3.214pt}{0.800pt}}
	\multiput(894.00,219.41)(2.971,0.506){31}{\rule{4.789pt}{0.122pt}}
	\multiput(894.00,216.34)(99.059,19.000){2}{\rule{2.395pt}{0.800pt}}
	\multiput(1003.00,238.41)(2.142,0.504){45}{\rule{3.554pt}{0.121pt}}
	\multiput(1003.00,235.34)(101.624,26.000){2}{\rule{1.777pt}{0.800pt}}
	\multiput(1112.00,264.41)(1.310,0.502){77}{\rule{2.276pt}{0.121pt}}
	\multiput(1112.00,261.34)(104.276,42.000){2}{\rule{1.138pt}{0.800pt}}
	\multiput(1221.00,306.41)(0.700,0.501){149}{\rule{1.318pt}{0.121pt}}
	\multiput(1221.00,303.34)(106.265,78.000){2}{\rule{0.659pt}{0.800pt}}
	\multiput(1331.41,383.00)(0.501,0.882){211}{\rule{0.121pt}{1.609pt}}
	\multiput(1328.34,383.00)(109.000,188.660){2}{\rule{0.800pt}{0.805pt}}
	\put(240,168){\makebox(0,0){$\ast$}}
	\put(349,173){\makebox(0,0){$\ast$}}
	\put(458,179){\makebox(0,0){$\ast$}}
	\put(567,186){\makebox(0,0){$\ast$}}
	\put(676,194){\makebox(0,0){$\ast$}}
	\put(785,204){\makebox(0,0){$\ast$}}
	\put(894,218){\makebox(0,0){$\ast$}}
	\put(1003,237){\makebox(0,0){$\ast$}}
	\put(1112,263){\makebox(0,0){$\ast$}}
	\put(1221,305){\makebox(0,0){$\ast$}}
	\put(1330,383){\makebox(0,0){$\ast$}}
	\put(1439,575){\makebox(0,0){$\ast$}}
	\put(601,735){\makebox(0,0){$\ast$}}
	\sbox{\plotpoint}{\rule[-0.200pt]{0.400pt}{0.400pt}}%
%	\put(131.0,131.0){\rule[-0.200pt]{0.400pt}{175.134pt}}
%	\put(131.0,131.0){\rule[-0.200pt]{315.097pt}{0.400pt}}
%	\put(1439.0,131.0){\rule[-0.200pt]{0.400pt}{175.134pt}}
%	\put(131.0,858.0){\rule[-0.200pt]{315.097pt}{0.400pt}}
\end{picture}
			\caption{ Comparison between three policies} \label{f3b} 
$\mu_1=5$, $\mu_2=2$, $q=0.2$
\end{center}
\end{figure}

Moving in the other direction from Figure \ref{f1}, $\mu_2$ and $q$ are 
both doubled: $\mu_2=2$, $q=0.2$, reducing the coefficient of variation. 
The resulting comparison, shown in Figure \ref{f3b}, is now fully 
predictable. The FCFS policy performs better, while the distance between 
LAS and FB-ph2 has increased.

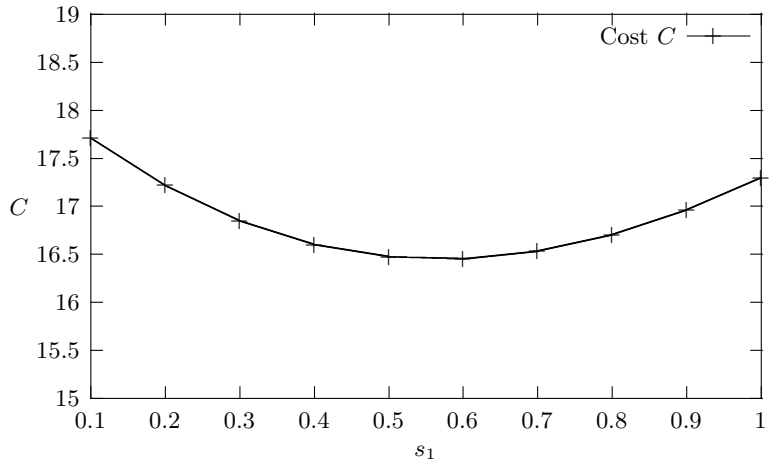
\begin{figure}[!ht]
	\begin{center}
% GNUPLOT: LaTeX picture
\setlength{\unitlength}{0.2pt}
\ifx\plotpoint\undefined\newsavebox{\plotpoint}\fi
\sbox{\plotpoint}{\rule[-0.200pt]{0.400pt}{0.400pt}}%
\begin{picture}(1500,900)(0,0)
\sbox{\plotpoint}{\rule[-0.200pt]{0.400pt}{0.400pt}}%
\put(171.0,131.0){\rule[-0.200pt]{4.818pt}{0.400pt}}
\put(151,131){\makebox(0,0)[r]{$15$}}
\put(1419.0,131.0){\rule[-0.200pt]{4.818pt}{0.400pt}}
\put(171.0,222.0){\rule[-0.200pt]{4.818pt}{0.400pt}}
\put(151,222){\makebox(0,0)[r]{$15.5$}}
\put(1419.0,222.0){\rule[-0.200pt]{4.818pt}{0.400pt}}
\put(171.0,313.0){\rule[-0.200pt]{4.818pt}{0.400pt}}
\put(151,313){\makebox(0,0)[r]{$16$}}
\put(1419.0,313.0){\rule[-0.200pt]{4.818pt}{0.400pt}}
\put(171.0,404.0){\rule[-0.200pt]{4.818pt}{0.400pt}}
\put(151,404){\makebox(0,0)[r]{$16.5$}}
\put(1419.0,404.0){\rule[-0.200pt]{4.818pt}{0.400pt}}
\put(171.0,495.0){\rule[-0.200pt]{4.818pt}{0.400pt}}
\put(151,495){\makebox(0,0)[r]{$17$}}
\put(1419.0,495.0){\rule[-0.200pt]{4.818pt}{0.400pt}}
\put(171.0,585.0){\rule[-0.200pt]{4.818pt}{0.400pt}}
\put(151,585){\makebox(0,0)[r]{$17.5$}}
\put(1419.0,585.0){\rule[-0.200pt]{4.818pt}{0.400pt}}
\put(171.0,676.0){\rule[-0.200pt]{4.818pt}{0.400pt}}
\put(151,676){\makebox(0,0)[r]{$18$}}
\put(1419.0,676.0){\rule[-0.200pt]{4.818pt}{0.400pt}}
\put(171.0,767.0){\rule[-0.200pt]{4.818pt}{0.400pt}}
\put(151,767){\makebox(0,0)[r]{$18.5$}}
\put(1419.0,767.0){\rule[-0.200pt]{4.818pt}{0.400pt}}
\put(171.0,858.0){\rule[-0.200pt]{4.818pt}{0.400pt}}
\put(151,858){\makebox(0,0)[r]{$19$}}
\put(1419.0,858.0){\rule[-0.200pt]{4.818pt}{0.400pt}}
\put(171.0,131.0){\rule[-0.200pt]{0.400pt}{4.818pt}}
\put(171,90){\makebox(0,0){$0.1$}}
\put(171.0,838.0){\rule[-0.200pt]{0.400pt}{4.818pt}}
\put(312.0,131.0){\rule[-0.200pt]{0.400pt}{4.818pt}}
\put(312,90){\makebox(0,0){$0.2$}}
\put(312.0,838.0){\rule[-0.200pt]{0.400pt}{4.818pt}}
\put(453.0,131.0){\rule[-0.200pt]{0.400pt}{4.818pt}}
\put(453,90){\makebox(0,0){$0.3$}}
\put(453.0,838.0){\rule[-0.200pt]{0.400pt}{4.818pt}}
\put(594.0,131.0){\rule[-0.200pt]{0.400pt}{4.818pt}}
\put(594,90){\makebox(0,0){$0.4$}}
\put(594.0,838.0){\rule[-0.200pt]{0.400pt}{4.818pt}}
\put(735.0,131.0){\rule[-0.200pt]{0.400pt}{4.818pt}}
\put(735,90){\makebox(0,0){$0.5$}}
\put(735.0,838.0){\rule[-0.200pt]{0.400pt}{4.818pt}}
\put(875.0,131.0){\rule[-0.200pt]{0.400pt}{4.818pt}}
\put(875,90){\makebox(0,0){$0.6$}}
\put(875.0,838.0){\rule[-0.200pt]{0.400pt}{4.818pt}}
\put(1016.0,131.0){\rule[-0.200pt]{0.400pt}{4.818pt}}
\put(1016,90){\makebox(0,0){$0.7$}}
\put(1016.0,838.0){\rule[-0.200pt]{0.400pt}{4.818pt}}
\put(1157.0,131.0){\rule[-0.200pt]{0.400pt}{4.818pt}}
\put(1157,90){\makebox(0,0){$0.8$}}
\put(1157.0,838.0){\rule[-0.200pt]{0.400pt}{4.818pt}}
\put(1298.0,131.0){\rule[-0.200pt]{0.400pt}{4.818pt}}
\put(1298,90){\makebox(0,0){$0.9$}}
\put(1298.0,838.0){\rule[-0.200pt]{0.400pt}{4.818pt}}
\put(1439.0,131.0){\rule[-0.200pt]{0.400pt}{4.818pt}}
\put(1439,90){\makebox(0,0){$1$}}
\put(1439.0,838.0){\rule[-0.200pt]{0.400pt}{4.818pt}}
\put(171.0,131.0){\rule[-0.200pt]{0.400pt}{146pt}}
\put(171.0,131.0){\rule[-0.200pt]{254pt}{0.400pt}}
\put(1439.0,131.0){\rule[-0.200pt]{0.400pt}{146pt}}
\put(171.0,858.0){\rule[-0.200pt]{254pt}{0.400pt}}
\put(36,494){\makebox(0,0){$C$}}
\put(805,29){\makebox(0,0){$s_1$}}
\put(1279,817){\makebox(0,0)[r]{Cost $C$}}
\put(1299.0,817.0){\rule[-0.200pt]{24.090pt}{0.400pt}}
\put(171,624){\usebox{\plotpoint}}
\multiput(171.00,622.92)(0.784,-0.499){177}{\rule{0.727pt}{0.120pt}}
\multiput(171.00,623.17)(139.492,-90.000){2}{\rule{0.363pt}{0.400pt}}
\multiput(312.00,532.92)(1.055,-0.499){131}{\rule{0.942pt}{0.120pt}}
\multiput(312.00,533.17)(139.045,-67.000){2}{\rule{0.471pt}{0.400pt}}
\multiput(453.00,465.92)(1.574,-0.498){87}{\rule{1.353pt}{0.120pt}}
\multiput(453.00,466.17)(138.191,-45.000){2}{\rule{0.677pt}{0.400pt}}
\multiput(594.00,420.92)(3.105,-0.496){43}{\rule{2.552pt}{0.120pt}}
\multiput(594.00,421.17)(135.703,-23.000){2}{\rule{1.276pt}{0.400pt}}
\multiput(735.00,397.94)(20.367,-0.468){5}{\rule{14.100pt}{0.113pt}}
\multiput(735.00,398.17)(110.735,-4.000){2}{\rule{7.050pt}{0.400pt}}
\multiput(875.00,395.58)(4.807,0.494){27}{\rule{3.860pt}{0.119pt}}
\multiput(875.00,394.17)(132.988,15.000){2}{\rule{1.930pt}{0.400pt}}
\multiput(1016.00,410.58)(2.294,0.497){59}{\rule{1.919pt}{0.120pt}}
\multiput(1016.00,409.17)(137.016,31.000){2}{\rule{0.960pt}{0.400pt}}
\multiput(1157.00,441.58)(1.507,0.498){91}{\rule{1.300pt}{0.120pt}}
\multiput(1157.00,440.17)(138.302,47.000){2}{\rule{0.650pt}{0.400pt}}
\multiput(1298.00,488.58)(1.159,0.499){119}{\rule{1.025pt}{0.120pt}}
\multiput(1298.00,487.17)(138.873,61.000){2}{\rule{0.512pt}{0.400pt}}
\put(171,624){\makebox(0,0){$+$}}
\put(312,534){\makebox(0,0){$+$}}
\put(453,467){\makebox(0,0){$+$}}
\put(594,422){\makebox(0,0){$+$}}
\put(735,399){\makebox(0,0){$+$}}
\put(875,395){\makebox(0,0){$+$}}
\put(1016,410){\makebox(0,0){$+$}}
\put(1157,441){\makebox(0,0){$+$}}
\put(1298,488){\makebox(0,0){$+$}}
\put(1439,549){\makebox(0,0){$+$}}
\put(1349,817){\makebox(0,0){$+$}}
\end{picture}
	\caption{ $K=2$: Average cost as function of $s_1$} \label{f2} 
$\lambda=2.5$, $\mu_1=5$, $\mu_2=1$, $q=0.1$, $c_1=1$, $c_2=20$
\end{center}
\end{figure}

The next experiment involves a system where $K=2$. There are three speed levels,
$s_0$ when the  processor is idle, $s_1$ when there is one job in the system and $s_2$ 
when 2 or more jobs are present. The following optimization problem arises in this 
context. Given a fixed idling speed $s_0$ and a fixed maximum speed $s_2$, how 
should one choose the intermediate speed $s_1$ in order to minimize the average cost? 
The trade-off here is between minimizing the holding costs, for which a high $s_1$ is 
better, and minimizing the energy costs, which requires a low $s_1$. That trade-off 
is illustrated in Figure \ref{f2}, where the cost function (\ref{C}) is plotted for $s_1$ 
varying between 0.1 and 1. The top service rates are $\mu_1=5$ and $\mu_2=1$, with 
$q=0.1$. The idling speed is assumed to be 0, and $s_1$ is treated as a fraction of 
the top speed. The arrival rate is fixed at $\lambda=2.5$ and the two cost coefficients 
are $c_1=1$ and $c_2=20$.

As expected, we observe a convex cost function. For this example, the 
optimal intermediate speed of the processor is 60\% of its maximal speed.

Next, we aim to quantify the benefits that can be derived by setting the 
intermediate speed levels optimally. In particular, one may ask  ``Would two 
intermediate speed levels produce significantly lower costs than one?'' We 
have carried out an experiment comparing three policies with the same fixed idling 
speed ($s_0=0$) and the same fixed top speed. These are: an unoptimized policy
without an intermediate speed level ($K=1$); a policy with one intermediate speed level 
chosen optimally ($K=2$, where the best value of $s_1$ is found by a search); and 
a policy with two intermediate speed levels chosen optimally ($K=3$, where the 
best values of $s_1$ and $s_2$ are found by a search). 

\begin{figure}[!ht]
	\begin{center}
% GNUPLOT: LaTeX picture
\setlength{\unitlength}{0.19pt}
\ifx\plotpoint\undefined\newsavebox{\plotpoint}\fi
\sbox{\plotpoint}{\rule[-0.200pt]{0.400pt}{0.400pt}}%
\begin{picture}(1500,900)(0,0)
\sbox{\plotpoint}{\rule[-0.200pt]{0.400pt}{0.400pt}}%
\put(131.0,131.0){\rule[-0.200pt]{4.818pt}{0.400pt}}
\put(111,131){\makebox(0,0)[r]{$0$}}
\put(1419.0,131.0){\rule[-0.200pt]{4.818pt}{0.400pt}}
\put(131.0,276.0){\rule[-0.200pt]{4.818pt}{0.400pt}}
\put(111,276){\makebox(0,0)[r]{$5$}}
\put(1419.0,276.0){\rule[-0.200pt]{4.818pt}{0.400pt}}
\put(131.0,422.0){\rule[-0.200pt]{4.818pt}{0.400pt}}
\put(111,422){\makebox(0,0)[r]{$10$}}
\put(1419.0,422.0){\rule[-0.200pt]{4.818pt}{0.400pt}}
\put(131.0,567.0){\rule[-0.200pt]{4.818pt}{0.400pt}}
\put(111,567){\makebox(0,0)[r]{$15$}}
\put(1419.0,567.0){\rule[-0.200pt]{4.818pt}{0.400pt}}
\put(131.0,713.0){\rule[-0.200pt]{4.818pt}{0.400pt}}
\put(111,713){\makebox(0,0)[r]{$20$}}
\put(1419.0,713.0){\rule[-0.200pt]{4.818pt}{0.400pt}}
\put(131.0,858.0){\rule[-0.200pt]{4.818pt}{0.400pt}}
\put(111,858){\makebox(0,0)[r]{$25$}}
\put(1419.0,858.0){\rule[-0.200pt]{4.818pt}{0.400pt}}
\put(131.0,131.0){\rule[-0.200pt]{0.400pt}{4.818pt}}
\put(131,90){\makebox(0,0){$0.5$}}
\put(131.0,838.0){\rule[-0.200pt]{0.400pt}{4.818pt}}
\put(393.0,131.0){\rule[-0.200pt]{0.400pt}{4.818pt}}
\put(393,90){\makebox(0,0){$1$}}
\put(393.0,838.0){\rule[-0.200pt]{0.400pt}{4.818pt}}
\put(654.0,131.0){\rule[-0.200pt]{0.400pt}{4.818pt}}
\put(654,90){\makebox(0,0){$1.5$}}
\put(654.0,838.0){\rule[-0.200pt]{0.400pt}{4.818pt}}
\put(916.0,131.0){\rule[-0.200pt]{0.400pt}{4.818pt}}
\put(916,90){\makebox(0,0){$2$}}
\put(916.0,838.0){\rule[-0.200pt]{0.400pt}{4.818pt}}
\put(1177.0,131.0){\rule[-0.200pt]{0.400pt}{4.818pt}}
\put(1177,90){\makebox(0,0){$2.5$}}
\put(1177.0,838.0){\rule[-0.200pt]{0.400pt}{4.818pt}}
\put(1439.0,131.0){\rule[-0.200pt]{0.400pt}{4.818pt}}
\put(1439,90){\makebox(0,0){$3$}}
\put(1439.0,838.0){\rule[-0.200pt]{0.400pt}{4.818pt}}
\put(131.0,131.0){\rule[-0.200pt]{0.400pt}{139pt}}
\put(131.0,131.0){\rule[-0.200pt]{250pt}{0.400pt}}
\put(1439.0,131.0){\rule[-0.200pt]{0.400pt}{139pt}}
\put(131.0,858.0){\rule[-0.200pt]{250pt}{0.400pt}}
\put(36,494){\makebox(0,0){$C$}}
\put(785,29){\makebox(0,0){$\lambda$}}
\put(511,817){\makebox(0,0)[r]{\small{Unoptimized}}}
\put(531.0,817.0){\rule[-0.200pt]{24.090pt}{0.400pt}}
\put(288,279){\usebox{\plotpoint}}
\multiput(288.00,279.58)(1.389,0.498){73}{\rule{1.205pt}{0.120pt}}
\multiput(288.00,278.17)(102.498,38.000){2}{\rule{0.603pt}{0.400pt}}
\multiput(393.00,317.58)(1.340,0.498){75}{\rule{1.167pt}{0.120pt}}
\multiput(393.00,316.17)(101.579,39.000){2}{\rule{0.583pt}{0.400pt}}
\multiput(497.00,356.58)(1.389,0.498){73}{\rule{1.205pt}{0.120pt}}
\multiput(497.00,355.17)(102.498,38.000){2}{\rule{0.603pt}{0.400pt}}
\multiput(602.00,394.58)(1.319,0.498){77}{\rule{1.150pt}{0.120pt}}
\multiput(602.00,393.17)(102.613,40.000){2}{\rule{0.575pt}{0.400pt}}
\multiput(707.00,434.58)(1.274,0.498){79}{\rule{1.115pt}{0.120pt}}
\multiput(707.00,433.17)(101.687,41.000){2}{\rule{0.557pt}{0.400pt}}
\multiput(811.00,475.58)(1.255,0.498){81}{\rule{1.100pt}{0.120pt}}
\multiput(811.00,474.17)(102.717,42.000){2}{\rule{0.550pt}{0.400pt}}
\multiput(916.00,517.58)(1.186,0.498){85}{\rule{1.045pt}{0.120pt}}
\multiput(916.00,516.17)(101.830,44.000){2}{\rule{0.523pt}{0.400pt}}
\multiput(1020.00,561.58)(1.121,0.498){91}{\rule{0.994pt}{0.120pt}}
\multiput(1020.00,560.17)(102.938,47.000){2}{\rule{0.497pt}{0.400pt}}
\multiput(1125.00,608.58)(0.975,0.498){105}{\rule{0.878pt}{0.120pt}}
\multiput(1125.00,607.17)(103.178,54.000){2}{\rule{0.439pt}{0.400pt}}
\multiput(1230.00,662.58)(0.801,0.499){127}{\rule{0.740pt}{0.120pt}}
\multiput(1230.00,661.17)(102.464,65.000){2}{\rule{0.370pt}{0.400pt}}
\multiput(1334.00,727.58)(0.535,0.499){193}{\rule{0.529pt}{0.120pt}}
\multiput(1334.00,726.17)(103.903,98.000){2}{\rule{0.264pt}{0.400pt}}
\put(288,279){\makebox(0,0){$+$}}
\put(393,317){\makebox(0,0){$+$}}
\put(497,356){\makebox(0,0){$+$}}
\put(602,394){\makebox(0,0){$+$}}
\put(707,434){\makebox(0,0){$+$}}
\put(811,475){\makebox(0,0){$+$}}
\put(916,517){\makebox(0,0){$+$}}
\put(1020,561){\makebox(0,0){$+$}}
\put(1125,608){\makebox(0,0){$+$}}
\put(1230,662){\makebox(0,0){$+$}}
\put(1334,727){\makebox(0,0){$+$}}
\put(1439,825){\makebox(0,0){$+$}}
\put(581,817){\makebox(0,0){$+$}}
\put(511,776){\makebox(0,0)[r]{\small{Optimized $K=2$}}}
\multiput(531,776)(20.756,0.000){5}{\usebox{\plotpoint}}
\put(631,776){\usebox{\plotpoint}}
\put(288,237){\usebox{\plotpoint}}
\multiput(288,237)(19.690,6.563){6}{\usebox{\plotpoint}}
\multiput(393,272)(19.555,6.957){5}{\usebox{\plotpoint}}
\multiput(497,309)(19.457,7.227){6}{\usebox{\plotpoint}}
\multiput(602,348)(19.334,7.549){5}{\usebox{\plotpoint}}
\multiput(707,389)(19.181,7.930){5}{\usebox{\plotpoint}}
\multiput(811,432)(19.011,8.329){6}{\usebox{\plotpoint}}
\multiput(916,478)(18.776,8.846){5}{\usebox{\plotpoint}}
\multiput(1020,527)(18.458,9.492){6}{\usebox{\plotpoint}}
\multiput(1125,581)(18.095,10.167){6}{\usebox{\plotpoint}}
\multiput(1230,640)(16.988,11.924){6}{\usebox{\plotpoint}}
\multiput(1334,713)(14.676,14.676){7}{\usebox{\plotpoint}}
\put(1439,818){\usebox{\plotpoint}}
\put(288,237){\makebox(0,0){$\times$}}
\put(393,272){\makebox(0,0){$\times$}}
\put(497,309){\makebox(0,0){$\times$}}
\put(602,348){\makebox(0,0){$\times$}}
\put(707,389){\makebox(0,0){$\times$}}
\put(811,432){\makebox(0,0){$\times$}}
\put(916,478){\makebox(0,0){$\times$}}
\put(1020,527){\makebox(0,0){$\times$}}
\put(1125,581){\makebox(0,0){$\times$}}
\put(1230,640){\makebox(0,0){$\times$}}
\put(1334,713){\makebox(0,0){$\times$}}
\put(1439,818){\makebox(0,0){$\times$}}
\put(581,776){\makebox(0,0){$\times$}}
\sbox{\plotpoint}{\rule[-0.400pt]{0.800pt}{0.800pt}}%
\sbox{\plotpoint}{\rule[-0.200pt]{0.400pt}{0.400pt}}%
\put(511,735){\makebox(0,0)[r]{\small{Optimized $K=3$}}}
\sbox{\plotpoint}{\rule[-0.400pt]{0.800pt}{0.800pt}}%
\put(531.0,735.0){\rule[-0.400pt]{24.090pt}{0.800pt}}
\put(288,224){\usebox{\plotpoint}}
\multiput(288.00,225.41)(1.614,0.503){59}{\rule{2.745pt}{0.121pt}}
\multiput(288.00,222.34)(99.302,33.000){2}{\rule{1.373pt}{0.800pt}}
\multiput(393.00,258.41)(1.505,0.503){63}{\rule{2.577pt}{0.121pt}}
\multiput(393.00,255.34)(98.651,35.000){2}{\rule{1.289pt}{0.800pt}}
\multiput(497.00,293.41)(1.360,0.503){71}{\rule{2.354pt}{0.121pt}}
\multiput(497.00,290.34)(100.114,39.000){2}{\rule{1.177pt}{0.800pt}}
\multiput(602.00,332.41)(1.293,0.502){75}{\rule{2.249pt}{0.121pt}}
\multiput(602.00,329.34)(100.333,41.000){2}{\rule{1.124pt}{0.800pt}}
\multiput(707.00,373.41)(1.192,0.502){81}{\rule{2.091pt}{0.121pt}}
\multiput(707.00,370.34)(99.660,44.000){2}{\rule{1.045pt}{0.800pt}}
\multiput(811.00,417.41)(1.101,0.502){89}{\rule{1.950pt}{0.121pt}}
\multiput(811.00,414.34)(100.953,48.000){2}{\rule{0.975pt}{0.800pt}}
\multiput(916.00,465.41)(1.026,0.502){95}{\rule{1.831pt}{0.121pt}}
\multiput(916.00,462.34)(100.199,51.000){2}{\rule{0.916pt}{0.800pt}}
\multiput(1020.00,516.41)(0.942,0.502){105}{\rule{1.700pt}{0.121pt}}
\multiput(1020.00,513.34)(101.472,56.000){2}{\rule{0.850pt}{0.800pt}}
\multiput(1125.00,572.41)(0.836,0.502){119}{\rule{1.533pt}{0.121pt}}
\multiput(1125.00,569.34)(101.817,63.000){2}{\rule{0.767pt}{0.800pt}}
\multiput(1230.00,635.41)(0.685,0.501){145}{\rule{1.295pt}{0.121pt}}
\multiput(1230.00,632.34)(101.313,76.000){2}{\rule{0.647pt}{0.800pt}}
\multiput(1335.41,710.00)(0.501,0.514){203}{\rule{0.121pt}{1.023pt}}
\multiput(1332.34,710.00)(105.000,105.877){2}{\rule{0.800pt}{0.511pt}}
\put(288,224){\makebox(0,0){$\ast$}}
\put(393,257){\makebox(0,0){$\ast$}}
\put(497,292){\makebox(0,0){$\ast$}}
\put(602,331){\makebox(0,0){$\ast$}}
\put(707,372){\makebox(0,0){$\ast$}}
\put(811,416){\makebox(0,0){$\ast$}}
\put(916,464){\makebox(0,0){$\ast$}}
\put(1020,515){\makebox(0,0){$\ast$}}
\put(1125,571){\makebox(0,0){$\ast$}}
\put(1230,634){\makebox(0,0){$\ast$}}
\put(1334,710){\makebox(0,0){$\ast$}}
\put(1439,818){\makebox(0,0){$\ast$}}
\put(581,735){\makebox(0,0){$\ast$}}
\sbox{\plotpoint}{\rule[-0.200pt]{0.400pt}{0.400pt}}%
%\put(131.0,131.0){\rule[-0.200pt]{0.400pt}{175.134pt}}
%\put(131.0,131.0){\rule[-0.200pt]{315.097pt}{0.400pt}}
%\put(1439.0,131.0){\rule[-0.200pt]{0.400pt}{175.134pt}}
%\put(131.0,858.0){\rule[-0.200pt]{315.097pt}{0.400pt}}
\end{picture}
	\caption{Unoptimized and optimized systems} \label{f3} 
$\mu_1=5$, $\mu_2=1$, $q=0.1$, $c_1=1$, $c_2=20$
\end{center}
\end{figure}
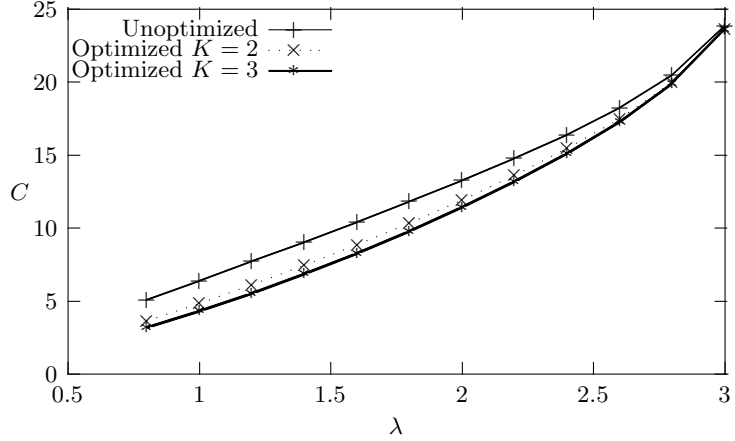

The reslts of this experiment are shown in Figure \ref{f3}, where the costs incurred 
by the three policies are plotted against the arrival rate $\lambda$. The parameters 
$\mu_1$, $\mu_2$ and $q$ are fixed, with the same values as before. The holding and 
energy cost coefficients are also kept at $c_1=1$, $c_2=20$.

Two aspects of the figure are worth pointing out. The first is the rather obvious observation that as the offered load increases, the advantage 
conferred by speed modulation gets smaller and eventually vanishes. That is 
because at high loads the cost function is dominated by the holding costs, 
and in order to minimize those, the processor must work at or near its 
maximum speed. The second observation is less obvious: the introduction of 
a second intermediary speed level ($K=3$) produces only a marginal improvement, 
compared to that achieved by one intermediary level. This phenomenon, which 
has been noted in other experiments, suggests that the case $K=2$ is the most 
important from a practical point of view.

The last question we wish to address concerns the possibility of using our 
2-phase FB model as an approximation for systems with more than two phases. 
Consider a 3-phase model with one foreground queue, $F$, and two background 
queues, $B_1$ and $B_2$. Incoming jobs arrive at rate $\lambda$ and join queue 
$F$. After completing phase 1, a job joins queue $B_1$ for phase 2 with probability 
$q_1$; after phase 2 it joins queue $B_2$ for phase 3 with probability $q_2$. Queue 
$B_1$ is served only when queue $F$ is empty, and queue $B_2$ is served only when 
both $F$ and $B_1$ are empty. Ignoring speed modulation, let the service rates in 
queues $F$, $B_1$ and $B_2$ be $\mu_1$, $\mu_2$ and $\mu_3$, respectively.

The first moment of the 3-phase service time distribution is 
\[
M_1 = \frac{1}{\mu_1} + \frac{q_1}{\mu_2} + \frac{q_1q_2}{\mu_3} \;.
\]
We hall attempt to approximate the 3-phase model using a 2-phase one with the 
same first moment. Moreover, the service rate in the foreground queue will be taken 
as $\mu_1$, and the probability of going to the background queue will be $q_1$. The 
service rate in the background queue, $\xi$, is chosen so that the first moment is 
the same:
\[
\frac{1}{\mu_1} + \frac{q_1}{\xi} = M_1\;.
\]
Hence,
\[
\frac{1}{\xi} = \frac{1}{\mu_2} + \frac{q_2}{\mu_3}\;.
\]

\begin{figure}[!ht]
\begin{center}
% GNUPLOT: LaTeX picture
\setlength{\unitlength}{0.19pt}
\ifx\plotpoint\undefined\newsavebox{\plotpoint}\fi
\sbox{\plotpoint}{\rule[-0.200pt]{0.400pt}{0.400pt}}%
\begin{picture}(1500,900)(0,0)
\sbox{\plotpoint}{\rule[-0.200pt]{0.400pt}{0.400pt}}%
\put(111.0,131.0){\rule[-0.200pt]{4.818pt}{0.400pt}}
\put(91,131){\makebox(0,0)[r]{$0$}}
\put(1419.0,131.0){\rule[-0.200pt]{4.818pt}{0.400pt}}
\put(111.0,212.0){\rule[-0.200pt]{4.818pt}{0.400pt}}
\put(91,212){\makebox(0,0)[r]{$1$}}
\put(1419.0,212.0){\rule[-0.200pt]{4.818pt}{0.400pt}}
\put(111.0,293.0){\rule[-0.200pt]{4.818pt}{0.400pt}}
\put(91,293){\makebox(0,0)[r]{$2$}}
\put(1419.0,293.0){\rule[-0.200pt]{4.818pt}{0.400pt}}
\put(111.0,373.0){\rule[-0.200pt]{4.818pt}{0.400pt}}
\put(91,373){\makebox(0,0)[r]{$3$}}
\put(1419.0,373.0){\rule[-0.200pt]{4.818pt}{0.400pt}}
\put(111.0,454.0){\rule[-0.200pt]{4.818pt}{0.400pt}}
\put(91,454){\makebox(0,0)[r]{$4$}}
\put(1419.0,454.0){\rule[-0.200pt]{4.818pt}{0.400pt}}
\put(111.0,535.0){\rule[-0.200pt]{4.818pt}{0.400pt}}
\put(91,535){\makebox(0,0)[r]{$5$}}
\put(1419.0,535.0){\rule[-0.200pt]{4.818pt}{0.400pt}}
\put(111.0,616.0){\rule[-0.200pt]{4.818pt}{0.400pt}}
\put(91,616){\makebox(0,0)[r]{$6$}}
\put(1419.0,616.0){\rule[-0.200pt]{4.818pt}{0.400pt}}
\put(111.0,696.0){\rule[-0.200pt]{4.818pt}{0.400pt}}
\put(91,696){\makebox(0,0)[r]{$7$}}
\put(1419.0,696.0){\rule[-0.200pt]{4.818pt}{0.400pt}}
\put(111.0,777.0){\rule[-0.200pt]{4.818pt}{0.400pt}}
\put(91,777){\makebox(0,0)[r]{$8$}}
\put(1419.0,777.0){\rule[-0.200pt]{4.818pt}{0.400pt}}
\put(111.0,858.0){\rule[-0.200pt]{4.818pt}{0.400pt}}
\put(91,858){\makebox(0,0)[r]{$9$}}
\put(1419.0,858.0){\rule[-0.200pt]{4.818pt}{0.400pt}}
\put(111.0,131.0){\rule[-0.200pt]{0.400pt}{4.818pt}}
\put(111,90){\makebox(0,0){$1.3$}}
\put(111.0,838.0){\rule[-0.200pt]{0.400pt}{4.818pt}}
\put(244.0,131.0){\rule[-0.200pt]{0.400pt}{4.818pt}}
\put(244,90){\makebox(0,0){$1.4$}}
\put(244.0,838.0){\rule[-0.200pt]{0.400pt}{4.818pt}}
\put(377.0,131.0){\rule[-0.200pt]{0.400pt}{4.818pt}}
\put(377,90){\makebox(0,0){$1.5$}}
\put(377.0,838.0){\rule[-0.200pt]{0.400pt}{4.818pt}}
\put(509.0,131.0){\rule[-0.200pt]{0.400pt}{4.818pt}}
\put(509,90){\makebox(0,0){$1.6$}}
\put(509.0,838.0){\rule[-0.200pt]{0.400pt}{4.818pt}}
\put(642.0,131.0){\rule[-0.200pt]{0.400pt}{4.818pt}}
\put(642,90){\makebox(0,0){$1.7$}}
\put(642.0,838.0){\rule[-0.200pt]{0.400pt}{4.818pt}}
\put(775.0,131.0){\rule[-0.200pt]{0.400pt}{4.818pt}}
\put(775,90){\makebox(0,0){$1.8$}}
\put(775.0,838.0){\rule[-0.200pt]{0.400pt}{4.818pt}}
\put(908.0,131.0){\rule[-0.200pt]{0.400pt}{4.818pt}}
\put(908,90){\makebox(0,0){$1.9$}}
\put(908.0,838.0){\rule[-0.200pt]{0.400pt}{4.818pt}}
\put(1041.0,131.0){\rule[-0.200pt]{0.400pt}{4.818pt}}
\put(1041,90){\makebox(0,0){$2$}}
\put(1041.0,838.0){\rule[-0.200pt]{0.400pt}{4.818pt}}
\put(1173.0,131.0){\rule[-0.200pt]{0.400pt}{4.818pt}}
\put(1173,90){\makebox(0,0){$2.1$}}
\put(1173.0,838.0){\rule[-0.200pt]{0.400pt}{4.818pt}}
\put(1306.0,131.0){\rule[-0.200pt]{0.400pt}{4.818pt}}
\put(1306,90){\makebox(0,0){$2.2$}}
\put(1306.0,838.0){\rule[-0.200pt]{0.400pt}{4.818pt}}
\put(1439.0,131.0){\rule[-0.200pt]{0.400pt}{4.818pt}}
\put(1439,90){\makebox(0,0){$2.3$}}
\put(1439.0,838.0){\rule[-0.200pt]{0.400pt}{4.818pt}}
\put(111.0,131.0){\rule[-0.200pt]{0.400pt}{139pt}}
\put(111.0,131.0){\rule[-0.200pt]{253pt}{0.400pt}}
\put(1439.0,131.0){\rule[-0.200pt]{0.400pt}{139pt}}
\put(111.0,858.0){\rule[-0.200pt]{253pt}{0.400pt}}
\put(36,494){\makebox(0,0){$L$}}
\put(775,29){\makebox(0,0){$\lambda$}}
\put(491,817){\makebox(0,0)[r]{\small{Approximation}}}
\put(511.0,817.0){\rule[-0.200pt]{24.090pt}{0.400pt}}
\put(244,218){\usebox{\plotpoint}}
\multiput(244.00,218.58)(5.254,0.493){23}{\rule{4.192pt}{0.119pt}}
\multiput(244.00,217.17)(124.299,13.000){2}{\rule{2.096pt}{0.400pt}}
\multiput(377.00,231.58)(3.956,0.495){31}{\rule{3.206pt}{0.119pt}}
\multiput(377.00,230.17)(125.346,17.000){2}{\rule{1.603pt}{0.400pt}}
\multiput(509.00,248.58)(3.558,0.495){35}{\rule{2.900pt}{0.119pt}}
\multiput(509.00,247.17)(126.981,19.000){2}{\rule{1.450pt}{0.400pt}}
\multiput(642.00,267.58)(2.690,0.497){47}{\rule{2.228pt}{0.120pt}}
\multiput(642.00,266.17)(128.376,25.000){2}{\rule{1.114pt}{0.400pt}}
\multiput(775.00,292.58)(2.095,0.497){61}{\rule{1.762pt}{0.120pt}}
\multiput(775.00,291.17)(129.342,32.000){2}{\rule{0.881pt}{0.400pt}}
\multiput(908.00,324.58)(1.519,0.498){85}{\rule{1.309pt}{0.120pt}}
\multiput(908.00,323.17)(130.283,44.000){2}{\rule{0.655pt}{0.400pt}}
\multiput(1041.00,368.58)(1.034,0.499){125}{\rule{0.925pt}{0.120pt}}
\multiput(1041.00,367.17)(130.080,64.000){2}{\rule{0.463pt}{0.400pt}}
\multiput(1173.00,432.58)(0.633,0.499){207}{\rule{0.607pt}{0.120pt}}
\multiput(1173.00,431.17)(131.741,105.000){2}{\rule{0.303pt}{0.400pt}}
\multiput(1306.58,537.00)(0.499,0.775){263}{\rule{0.120pt}{0.720pt}}
\multiput(1305.17,537.00)(133.000,204.507){2}{\rule{0.400pt}{0.360pt}}
\put(244,218){\makebox(0,0){$+$}}
\put(377,231){\makebox(0,0){$+$}}
\put(509,248){\makebox(0,0){$+$}}
\put(642,267){\makebox(0,0){$+$}}
\put(775,292){\makebox(0,0){$+$}}
\put(908,324){\makebox(0,0){$+$}}
\put(1041,368){\makebox(0,0){$+$}}
\put(1173,432){\makebox(0,0){$+$}}
\put(1306,537){\makebox(0,0){$+$}}
\put(1439,743){\makebox(0,0){$+$}}
\put(561,817){\makebox(0,0){$+$}}
\put(491,776){\makebox(0,0)[r]{\small{Simulation}}}
\multiput(511,776)(20.756,0.000){5}{\usebox{\plotpoint}}
\put(611,776){\usebox{\plotpoint}}
\put(244,218){\usebox{\plotpoint}}
\multiput(244,218)(20.657,2.019){7}{\usebox{\plotpoint}}
\multiput(377,231)(20.585,2.651){6}{\usebox{\plotpoint}}
\multiput(509,248)(20.588,2.632){7}{\usebox{\plotpoint}}
\multiput(642,265)(20.426,3.686){6}{\usebox{\plotpoint}}
\multiput(775,289)(19.996,5.563){7}{\usebox{\plotpoint}}
\multiput(908,326)(19.792,6.250){7}{\usebox{\plotpoint}}
\multiput(1041,368)(18.279,9.832){7}{\usebox{\plotpoint}}
\multiput(1173,439)(16.530,12.553){8}{\usebox{\plotpoint}}
\multiput(1306,540)(11.258,17.437){12}{\usebox{\plotpoint}}
\put(1439,746){\usebox{\plotpoint}}
\put(244,218){\makebox(0,0){$\times$}}
\put(377,231){\makebox(0,0){$\times$}}
\put(509,248){\makebox(0,0){$\times$}}
\put(642,265){\makebox(0,0){$\times$}}
\put(775,289){\makebox(0,0){$\times$}}
\put(908,326){\makebox(0,0){$\times$}}
\put(1041,368){\makebox(0,0){$\times$}}
\put(1173,439){\makebox(0,0){$\times$}}
\put(1306,540){\makebox(0,0){$\times$}}
\put(1439,746){\makebox(0,0){$\times$}}
\put(561,776){\makebox(0,0){$\times$}}
\end{picture}
	\caption{Three phases approximated by two phases} \label{f4} 
	$\mu_1=5$, $\mu_2=1$, $\mu_3=0.5$; $q_1=0.1$, $q_2=0.5$
\end{center}
\end{figure}
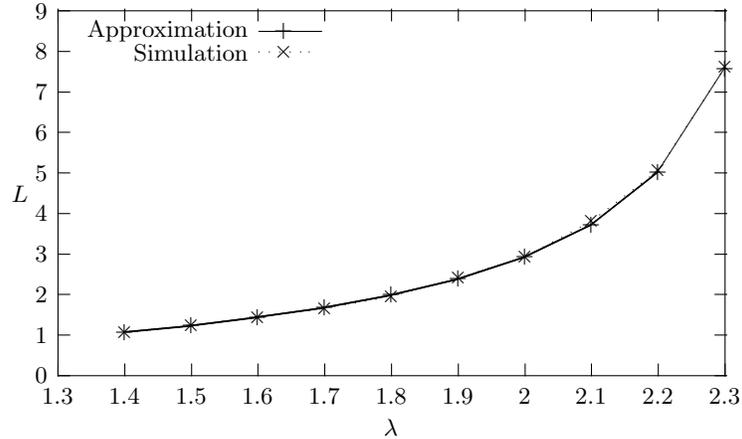

We have taken an example 3-phase model with parameters $\mu_1=5$, $\mu_2=1$, 
$\mu_3=0.5$; $q_1=0.1$, $q_2=0.5$. That is, phase 3 is twice as long as phase 2, on 
the average, and a job is five times more likely to move from phase 2 to phase 3, than 
from phase 1 to phase 2. In the absence of an exact solution, that model is simulated.
The approximating 2-phase model is evaluated using the $K=1$ solution. In Figure \ref{f4}, 
the average total number of jobs present, $L$, in the the 2-phase and 3-phase systems, 
is plotted against the arrival rate, $\lambda$. Each simulated point is the result of a run 
during which one million jobs arrive into the system.

We observe that the 2-phase model provides an excellent approximation to the 
3-phase one. The two plots are almost indistinguishible. However, perhaps this 
example is too favourable. Only about one job in twenty reaches queue $B_2$. 
In order to construct a more stringent test we have repeated the exercise with the 
following parameters:  $\mu_1=5$, $\mu_2=3$, $\mu_3=3$; $q_1=0.6$, $q_2=0.8$. 
This is a system where the three phases have comparable lengths and almost half of all jobs reach queue $B_2$. The stability condition imposes a lower bound on the arrival rate.

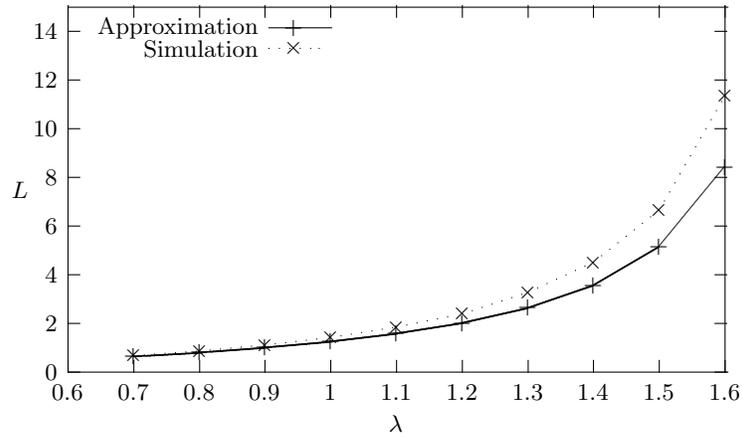
\begin{figure}[!ht]
	\begin{center}
% GNUPLOT: LaTeX picture
\setlength{\unitlength}{0.19pt}
\ifx\plotpoint\undefined\newsavebox{\plotpoint}\fi
\sbox{\plotpoint}{\rule[-0.200pt]{0.400pt}{0.400pt}}%
\begin{picture}(1500,900)(0,0)
\sbox{\plotpoint}{\rule[-0.200pt]{0.400pt}{0.400pt}}%
\put(131.0,131.0){\rule[-0.200pt]{4.818pt}{0.400pt}}
\put(111,131){\makebox(0,0)[r]{$0$}}
\put(1419.0,131.0){\rule[-0.200pt]{4.818pt}{0.400pt}}
\put(131.0,228.0){\rule[-0.200pt]{4.818pt}{0.400pt}}
\put(111,228){\makebox(0,0)[r]{$2$}}
\put(1419.0,228.0){\rule[-0.200pt]{4.818pt}{0.400pt}}
\put(131.0,325.0){\rule[-0.200pt]{4.818pt}{0.400pt}}
\put(111,325){\makebox(0,0)[r]{$4$}}
\put(1419.0,325.0){\rule[-0.200pt]{4.818pt}{0.400pt}}
\put(131.0,422.0){\rule[-0.200pt]{4.818pt}{0.400pt}}
\put(111,422){\makebox(0,0)[r]{$6$}}
\put(1419.0,422.0){\rule[-0.200pt]{4.818pt}{0.400pt}}
\put(131.0,519.0){\rule[-0.200pt]{4.818pt}{0.400pt}}
\put(111,519){\makebox(0,0)[r]{$8$}}
\put(1419.0,519.0){\rule[-0.200pt]{4.818pt}{0.400pt}}
\put(131.0,616.0){\rule[-0.200pt]{4.818pt}{0.400pt}}
\put(111,616){\makebox(0,0)[r]{$10$}}
\put(1419.0,616.0){\rule[-0.200pt]{4.818pt}{0.400pt}}
\put(131.0,713.0){\rule[-0.200pt]{4.818pt}{0.400pt}}
\put(111,713){\makebox(0,0)[r]{$12$}}
\put(1419.0,713.0){\rule[-0.200pt]{4.818pt}{0.400pt}}
\put(131.0,810.0){\rule[-0.200pt]{4.818pt}{0.400pt}}
\put(111,810){\makebox(0,0)[r]{$14$}}
\put(1419.0,810.0){\rule[-0.200pt]{4.818pt}{0.400pt}}
\put(131.0,131.0){\rule[-0.200pt]{0.400pt}{4.818pt}}
\put(131,90){\makebox(0,0){$0.6$}}
\put(131.0,838.0){\rule[-0.200pt]{0.400pt}{4.818pt}}
\put(262.0,131.0){\rule[-0.200pt]{0.400pt}{4.818pt}}
\put(262,90){\makebox(0,0){$0.7$}}
\put(262.0,838.0){\rule[-0.200pt]{0.400pt}{4.818pt}}
\put(393.0,131.0){\rule[-0.200pt]{0.400pt}{4.818pt}}
\put(393,90){\makebox(0,0){$0.8$}}
\put(393.0,838.0){\rule[-0.200pt]{0.400pt}{4.818pt}}
\put(523.0,131.0){\rule[-0.200pt]{0.400pt}{4.818pt}}
\put(523,90){\makebox(0,0){$0.9$}}
\put(523.0,838.0){\rule[-0.200pt]{0.400pt}{4.818pt}}
\put(654.0,131.0){\rule[-0.200pt]{0.400pt}{4.818pt}}
\put(654,90){\makebox(0,0){$1$}}
\put(654.0,838.0){\rule[-0.200pt]{0.400pt}{4.818pt}}
\put(785.0,131.0){\rule[-0.200pt]{0.400pt}{4.818pt}}
\put(785,90){\makebox(0,0){$1.1$}}
\put(785.0,838.0){\rule[-0.200pt]{0.400pt}{4.818pt}}
\put(916.0,131.0){\rule[-0.200pt]{0.400pt}{4.818pt}}
\put(916,90){\makebox(0,0){$1.2$}}
\put(916.0,838.0){\rule[-0.200pt]{0.400pt}{4.818pt}}
\put(1047.0,131.0){\rule[-0.200pt]{0.400pt}{4.818pt}}
\put(1047,90){\makebox(0,0){$1.3$}}
\put(1047.0,838.0){\rule[-0.200pt]{0.400pt}{4.818pt}}
\put(1177.0,131.0){\rule[-0.200pt]{0.400pt}{4.818pt}}
\put(1177,90){\makebox(0,0){$1.4$}}
\put(1177.0,838.0){\rule[-0.200pt]{0.400pt}{4.818pt}}
\put(1308.0,131.0){\rule[-0.200pt]{0.400pt}{4.818pt}}
\put(1308,90){\makebox(0,0){$1.5$}}
\put(1308.0,838.0){\rule[-0.200pt]{0.400pt}{4.818pt}}
\put(1439.0,131.0){\rule[-0.200pt]{0.400pt}{4.818pt}}
\put(1439,90){\makebox(0,0){$1.6$}}
\put(1439.0,838.0){\rule[-0.200pt]{0.400pt}{4.818pt}}
\put(131.0,131.0){\rule[-0.200pt]{0.400pt}{139pt}}
\put(131.0,131.0){\rule[-0.200pt]{250pt}{0.400pt}}
\put(1439.0,131.0){\rule[-0.200pt]{0.400pt}{139pt}}
\put(131.0,858.0){\rule[-0.200pt]{250pt}{0.400pt}}
\put(36,494){\makebox(0,0){$L$}}
\put(785,29){\makebox(0,0){$\lambda$}}
\put(511,817){\makebox(0,0)[r]{\small{Approximation}}}
\put(531.0,817.0){\rule[-0.200pt]{24.090pt}{0.400pt}}
\put(262,162){\usebox{\plotpoint}}
\multiput(262.00,162.59)(8.616,0.488){13}{\rule{6.650pt}{0.117pt}}
\multiput(262.00,161.17)(117.198,8.000){2}{\rule{3.325pt}{0.400pt}}
\multiput(393.00,170.58)(6.750,0.491){17}{\rule{5.300pt}{0.118pt}}
\multiput(393.00,169.17)(119.000,10.000){2}{\rule{2.650pt}{0.400pt}}
\multiput(523.00,180.58)(5.623,0.492){21}{\rule{4.467pt}{0.119pt}}
\multiput(523.00,179.17)(121.729,12.000){2}{\rule{2.233pt}{0.400pt}}
\multiput(654.00,192.58)(4.178,0.494){29}{\rule{3.375pt}{0.119pt}}
\multiput(654.00,191.17)(123.995,16.000){2}{\rule{1.688pt}{0.400pt}}
\multiput(785.00,208.58)(3.164,0.496){39}{\rule{2.595pt}{0.119pt}}
\multiput(785.00,207.17)(125.613,21.000){2}{\rule{1.298pt}{0.400pt}}
\multiput(916.00,229.58)(2.202,0.497){57}{\rule{1.847pt}{0.120pt}}
\multiput(916.00,228.17)(127.167,30.000){2}{\rule{0.923pt}{0.400pt}}
\multiput(1047.00,259.58)(1.451,0.498){87}{\rule{1.256pt}{0.120pt}}
\multiput(1047.00,258.17)(127.394,45.000){2}{\rule{0.628pt}{0.400pt}}
\multiput(1177.00,304.58)(0.852,0.499){151}{\rule{0.781pt}{0.120pt}}
\multiput(1177.00,303.17)(129.380,77.000){2}{\rule{0.390pt}{0.400pt}}
\multiput(1308.58,381.00)(0.499,0.607){259}{\rule{0.120pt}{0.585pt}}
\multiput(1307.17,381.00)(131.000,157.785){2}{\rule{0.400pt}{0.293pt}}
\put(262,162){\makebox(0,0){$+$}}
\put(393,170){\makebox(0,0){$+$}}
\put(523,180){\makebox(0,0){$+$}}
\put(654,192){\makebox(0,0){$+$}}
\put(785,208){\makebox(0,0){$+$}}
\put(916,229){\makebox(0,0){$+$}}
\put(1047,259){\makebox(0,0){$+$}}
\put(1177,304){\makebox(0,0){$+$}}
\put(1308,381){\makebox(0,0){$+$}}
\put(1439,540){\makebox(0,0){$+$}}
\put(581,817){\makebox(0,0){$+$}}
\put(511,776){\makebox(0,0)[r]{\small{Simulation}}}
\multiput(531,776)(20.756,0.000){5}{\usebox{\plotpoint}}
\put(631,776){\usebox{\plotpoint}}
\put(262,164){\usebox{\plotpoint}}
\multiput(262,164)(20.707,1.423){7}{\usebox{\plotpoint}}
\multiput(393,173)(20.668,1.908){6}{\usebox{\plotpoint}}
\multiput(523,185)(20.621,2.361){6}{\usebox{\plotpoint}}
\multiput(654,200)(20.494,3.285){7}{\usebox{\plotpoint}}
\multiput(785,221)(20.358,4.041){6}{\usebox{\plotpoint}}
\multiput(916,247)(19.720,6.473){7}{\usebox{\plotpoint}}
\multiput(1047,290)(18.845,8.698){7}{\usebox{\plotpoint}}
\multiput(1177,350)(16.256,12.905){8}{\usebox{\plotpoint}}
\multiput(1308,454)(10.340,17.996){13}{\usebox{\plotpoint}}
\put(1439,682){\usebox{\plotpoint}}
\put(262,164){\makebox(0,0){$\times$}}
\put(393,173){\makebox(0,0){$\times$}}
\put(523,185){\makebox(0,0){$\times$}}
\put(654,200){\makebox(0,0){$\times$}}
\put(785,221){\makebox(0,0){$\times$}}
\put(916,247){\makebox(0,0){$\times$}}
\put(1047,290){\makebox(0,0){$\times$}}
\put(1177,350){\makebox(0,0){$\times$}}
\put(1308,454){\makebox(0,0){$\times$}}
\put(1439,682){\makebox(0,0){$\times$}}
\put(581,776){\makebox(0,0){$\times$}}
%\put(131.0,131.0){\rule[-0.200pt]{0.400pt}{175.134pt}}
%\put(131.0,131.0){\rule[-0.200pt]{315.097pt}{0.400pt}}
%\put(1439.0,131.0){\rule[-0.200pt]{0.400pt}{175.134pt}}
%\put(131.0,858.0){\rule[-0.200pt]{315.097pt}{0.400pt}}
\end{picture}
\caption{Three phases approximated by two phases} \label{f5} 
$\mu_1=5$, $\mu_2=3$, $\mu_3=3$; $q_1=0.6$, $q_2=0.8$, $c_1=1$
\end{center}
\end{figure}

The new results are shown in Figure \ref{f5}.

This time the approximation noticeably underestimates $L$, 
particularly when the load is very heavy. At light and moderate 
loads, the approximation is still acceptable. 

In the final experiment, the results of the previous section are applied 
to a capacity-modulated multiprocessor system with 10 servers. 
The aim is to examine the effect of the threshold parameter, $K$, on 
the cost function (\ref{Cm}).

\begin{figure}[!ht]
	\begin{center}
% GNUPLOT: LaTeX picture
\setlength{\unitlength}{0.19pt}
\ifx\plotpoint\undefined\newsavebox{\plotpoint}\fi
\sbox{\plotpoint}{\rule[-0.200pt]{0.400pt}{0.400pt}}%
\begin{picture}(1500,900)(0,0)
	\sbox{\plotpoint}{\rule[-0.200pt]{0.400pt}{0.400pt}}%
	\put(131.0,131.0){\rule[-0.200pt]{4.818pt}{0.400pt}}
	\put(111,131){\makebox(0,0)[r]{$10$}}
	\put(1419.0,131.0){\rule[-0.200pt]{4.818pt}{0.400pt}}
	\put(131.0,212.0){\rule[-0.200pt]{4.818pt}{0.400pt}}
	\put(111,212){\makebox(0,0)[r]{$12$}}
	\put(1419.0,212.0){\rule[-0.200pt]{4.818pt}{0.400pt}}
	\put(131.0,293.0){\rule[-0.200pt]{4.818pt}{0.400pt}}
	\put(111,293){\makebox(0,0)[r]{$14$}}
	\put(1419.0,293.0){\rule[-0.200pt]{4.818pt}{0.400pt}}
	\put(131.0,373.0){\rule[-0.200pt]{4.818pt}{0.400pt}}
	\put(111,373){\makebox(0,0)[r]{$16$}}
	\put(1419.0,373.0){\rule[-0.200pt]{4.818pt}{0.400pt}}
	\put(131.0,454.0){\rule[-0.200pt]{4.818pt}{0.400pt}}
	\put(111,454){\makebox(0,0)[r]{$18$}}
	\put(1419.0,454.0){\rule[-0.200pt]{4.818pt}{0.400pt}}
	\put(131.0,535.0){\rule[-0.200pt]{4.818pt}{0.400pt}}
	\put(111,535){\makebox(0,0)[r]{$20$}}
	\put(1419.0,535.0){\rule[-0.200pt]{4.818pt}{0.400pt}}
	\put(131.0,616.0){\rule[-0.200pt]{4.818pt}{0.400pt}}
	\put(111,616){\makebox(0,0)[r]{$22$}}
	\put(1419.0,616.0){\rule[-0.200pt]{4.818pt}{0.400pt}}
	\put(131.0,696.0){\rule[-0.200pt]{4.818pt}{0.400pt}}
	\put(111,696){\makebox(0,0)[r]{$24$}}
	\put(1419.0,696.0){\rule[-0.200pt]{4.818pt}{0.400pt}}
	\put(131.0,777.0){\rule[-0.200pt]{4.818pt}{0.400pt}}
	\put(111,777){\makebox(0,0)[r]{$26$}}
	\put(1419.0,777.0){\rule[-0.200pt]{4.818pt}{0.400pt}}
	\put(131.0,858.0){\rule[-0.200pt]{4.818pt}{0.400pt}}
	\put(111,858){\makebox(0,0)[r]{$28$}}
	\put(1419.0,858.0){\rule[-0.200pt]{4.818pt}{0.400pt}}
	\put(131.0,131.0){\rule[-0.200pt]{0.400pt}{4.818pt}}
	\put(131,90){\makebox(0,0){$0$}}
	\put(131.0,838.0){\rule[-0.200pt]{0.400pt}{4.818pt}}
	\put(276.0,131.0){\rule[-0.200pt]{0.400pt}{4.818pt}}
	\put(276,90){\makebox(0,0){$1$}}
	\put(276.0,838.0){\rule[-0.200pt]{0.400pt}{4.818pt}}
	\put(422.0,131.0){\rule[-0.200pt]{0.400pt}{4.818pt}}
	\put(422,90){\makebox(0,0){$2$}}
	\put(422.0,838.0){\rule[-0.200pt]{0.400pt}{4.818pt}}
	\put(567.0,131.0){\rule[-0.200pt]{0.400pt}{4.818pt}}
	\put(567,90){\makebox(0,0){$3$}}
	\put(567.0,838.0){\rule[-0.200pt]{0.400pt}{4.818pt}}
	\put(712.0,131.0){\rule[-0.200pt]{0.400pt}{4.818pt}}
	\put(712,90){\makebox(0,0){$4$}}
	\put(712.0,838.0){\rule[-0.200pt]{0.400pt}{4.818pt}}
	\put(858.0,131.0){\rule[-0.200pt]{0.400pt}{4.818pt}}
	\put(858,90){\makebox(0,0){$5$}}
	\put(858.0,838.0){\rule[-0.200pt]{0.400pt}{4.818pt}}
	\put(1003.0,131.0){\rule[-0.200pt]{0.400pt}{4.818pt}}
	\put(1003,90){\makebox(0,0){$6$}}
	\put(1003.0,838.0){\rule[-0.200pt]{0.400pt}{4.818pt}}
	\put(1148.0,131.0){\rule[-0.200pt]{0.400pt}{4.818pt}}
	\put(1148,90){\makebox(0,0){$7$}}
	\put(1148.0,838.0){\rule[-0.200pt]{0.400pt}{4.818pt}}
	\put(1294.0,131.0){\rule[-0.200pt]{0.400pt}{4.818pt}}
	\put(1294,90){\makebox(0,0){$8$}}
	\put(1294.0,838.0){\rule[-0.200pt]{0.400pt}{4.818pt}}
	\put(1439.0,131.0){\rule[-0.200pt]{0.400pt}{4.818pt}}
	\put(1439,90){\makebox(0,0){$9$}}
	\put(1439.0,838.0){\rule[-0.200pt]{0.400pt}{4.818pt}}
\put(131.0,131.0){\rule[-0.200pt]{0.400pt}{139pt}}
\put(131.0,131.0){\rule[-0.200pt]{250pt}{0.400pt}}
\put(1439.0,131.0){\rule[-0.200pt]{0.400pt}{139pt}}
\put(131.0,858.0){\rule[-0.200pt]{250pt}{0.400pt}}
	\put(36,494){\makebox(0,0){$C$}}
	\put(785,29){\makebox(0,0){$K$}}
	\put(591,817){\makebox(0,0)[r]{$c_2=0.5$}}
	\put(611.0,817.0){\rule[-0.200pt]{24.090pt}{0.400pt}}
	\put(131,257){\usebox{\plotpoint}}
	\put(131,255.67){\rule{34.931pt}{0.400pt}}
	\multiput(131.00,256.17)(72.500,-1.000){2}{\rule{17.465pt}{0.400pt}}
	\put(276,254.67){\rule{35.171pt}{0.400pt}}
	\multiput(276.00,255.17)(73.000,-1.000){2}{\rule{17.586pt}{0.400pt}}
	\put(422,253.67){\rule{34.931pt}{0.400pt}}
	\multiput(422.00,254.17)(72.500,-1.000){2}{\rule{17.465pt}{0.400pt}}
	\put(567,253.67){\rule{34.931pt}{0.400pt}}
	\multiput(567.00,253.17)(72.500,1.000){2}{\rule{17.465pt}{0.400pt}}
	\multiput(712.00,255.59)(16.183,0.477){7}{\rule{11.780pt}{0.115pt}}
	\multiput(712.00,254.17)(121.550,5.000){2}{\rule{5.890pt}{0.400pt}}
	\multiput(858.00,260.58)(7.532,0.491){17}{\rule{5.900pt}{0.118pt}}
	\multiput(858.00,259.17)(132.754,10.000){2}{\rule{2.950pt}{0.400pt}}
	\multiput(1003.00,270.58)(4.626,0.494){29}{\rule{3.725pt}{0.119pt}}
	\multiput(1003.00,269.17)(137.269,16.000){2}{\rule{1.863pt}{0.400pt}}
	\multiput(1148.00,286.58)(3.215,0.496){43}{\rule{2.639pt}{0.120pt}}
	\multiput(1148.00,285.17)(140.522,23.000){2}{\rule{1.320pt}{0.400pt}}
	\multiput(1294.00,309.58)(2.359,0.497){59}{\rule{1.971pt}{0.120pt}}
	\multiput(1294.00,308.17)(140.909,31.000){2}{\rule{0.985pt}{0.400pt}}
	\put(131,257){\makebox(0,0){$+$}}
	\put(276,256){\makebox(0,0){$+$}}
	\put(422,255){\makebox(0,0){$+$}}
	\put(567,254){\makebox(0,0){$+$}}
	\put(712,255){\makebox(0,0){$+$}}
	\put(858,260){\makebox(0,0){$+$}}
	\put(1003,270){\makebox(0,0){$+$}}
	\put(1148,286){\makebox(0,0){$+$}}
	\put(1294,309){\makebox(0,0){$+$}}
	\put(1439,340){\makebox(0,0){$+$}}
	\put(661,817){\makebox(0,0){$+$}}
	\put(591,776){\makebox(0,0)[r]{$c_2=1$}}
	\multiput(611,776)(20.756,0.000){5}{\usebox{\plotpoint}}
	\put(711,776){\usebox{\plotpoint}}
	\put(131,458){\usebox{\plotpoint}}
	\multiput(131,458)(20.755,-0.143){7}{\usebox{\plotpoint}}
	\multiput(276,457)(20.751,-0.426){8}{\usebox{\plotpoint}}
	\multiput(422,454)(20.743,-0.715){7}{\usebox{\plotpoint}}
	\multiput(567,449)(20.743,-0.715){7}{\usebox{\plotpoint}}
	\multiput(712,444)(20.748,-0.568){7}{\usebox{\plotpoint}}
	\multiput(858,440)(20.755,0.143){7}{\usebox{\plotpoint}}
	\multiput(1003,441)(20.731,1.001){7}{\usebox{\plotpoint}}
	\multiput(1148,448)(20.632,2.261){7}{\usebox{\plotpoint}}
	\multiput(1294,464)(20.405,3.800){7}{\usebox{\plotpoint}}
	\put(1439,491){\usebox{\plotpoint}}
	\put(131,458){\makebox(0,0){$\times$}}
	\put(276,457){\makebox(0,0){$\times$}}
	\put(422,454){\makebox(0,0){$\times$}}
	\put(567,449){\makebox(0,0){$\times$}}
	\put(712,444){\makebox(0,0){$\times$}}
	\put(858,440){\makebox(0,0){$\times$}}
	\put(1003,441){\makebox(0,0){$\times$}}
	\put(1148,448){\makebox(0,0){$\times$}}
	\put(1294,464){\makebox(0,0){$\times$}}
	\put(1439,491){\makebox(0,0){$\times$}}
	\put(661,776){\makebox(0,0){$\times$}}
	\sbox{\plotpoint}{\rule[-0.400pt]{0.800pt}{0.800pt}}%
	\sbox{\plotpoint}{\rule[-0.200pt]{0.400pt}{0.400pt}}%
	\put(591,735){\makebox(0,0)[r]{$c_2=1.5$}}
	\sbox{\plotpoint}{\rule[-0.400pt]{0.800pt}{0.800pt}}%
	\put(611.0,735.0){\rule[-0.400pt]{24.090pt}{0.800pt}}
	\put(131,660){\usebox{\plotpoint}}
	\put(131,657.84){\rule{34.931pt}{0.800pt}}
	\multiput(131.00,658.34)(72.500,-1.000){2}{\rule{17.465pt}{0.800pt}}
	\multiput(276.00,657.06)(24.100,-0.560){3}{\rule{23.560pt}{0.135pt}}
	\multiput(276.00,657.34)(97.100,-5.000){2}{\rule{11.780pt}{0.800pt}}
	\multiput(422.00,652.08)(8.001,-0.514){13}{\rule{11.800pt}{0.124pt}}
	\multiput(422.00,652.34)(120.509,-10.000){2}{\rule{5.900pt}{0.800pt}}
	\multiput(567.00,642.08)(6.503,-0.511){17}{\rule{9.867pt}{0.123pt}}
	\multiput(567.00,642.34)(124.521,-12.000){2}{\rule{4.933pt}{0.800pt}}
	\multiput(712.00,630.08)(6.548,-0.511){17}{\rule{9.933pt}{0.123pt}}
	\multiput(712.00,630.34)(125.383,-12.000){2}{\rule{4.967pt}{0.800pt}}
	\multiput(858.00,618.08)(10.497,-0.520){9}{\rule{14.700pt}{0.125pt}}
	\multiput(858.00,618.34)(114.489,-8.000){2}{\rule{7.350pt}{0.800pt}}
	\put(1003,609.84){\rule{34.931pt}{0.800pt}}
	\multiput(1003.00,610.34)(72.500,-1.000){2}{\rule{17.465pt}{0.800pt}}
	\multiput(1148.00,612.40)(9.130,0.516){11}{\rule{13.178pt}{0.124pt}}
	\multiput(1148.00,609.34)(118.649,9.000){2}{\rule{6.589pt}{0.800pt}}
	\multiput(1294.00,621.41)(3.243,0.505){39}{\rule{5.243pt}{0.122pt}}
	\multiput(1294.00,618.34)(134.117,23.000){2}{\rule{2.622pt}{0.800pt}}
	\put(131,660){\makebox(0,0){$\ast$}}
	\put(276,659){\makebox(0,0){$\ast$}}
	\put(422,654){\makebox(0,0){$\ast$}}
	\put(567,644){\makebox(0,0){$\ast$}}
	\put(712,632){\makebox(0,0){$\ast$}}
	\put(858,620){\makebox(0,0){$\ast$}}
	\put(1003,612){\makebox(0,0){$\ast$}}
	\put(1148,611){\makebox(0,0){$\ast$}}
	\put(1294,620){\makebox(0,0){$\ast$}}
	\put(1439,643){\makebox(0,0){$\ast$}}
	\put(661,735){\makebox(0,0){$\ast$}}
	\sbox{\plotpoint}{\rule[-0.200pt]{0.400pt}{0.400pt}}%
%	\put(131.0,131.0){\rule[-0.200pt]{0.400pt}{175.134pt}}
%	\put(131.0,131.0){\rule[-0.200pt]{315.097pt}{0.400pt}}
%	\put(1439.0,131.0){\rule[-0.200pt]{0.400pt}{175.134pt}}
%	\put(131.0,858.0){\rule[-0.200pt]{315.097pt}{0.400pt}}
\end{picture}
\caption{Capacity-modulated multiprocessor system} \label{f6} 
$m=10$, $\lambda =5$, $\mu_1=1$, $\mu_2=0.2$, $q=0.1$
\end{center}
\end{figure}
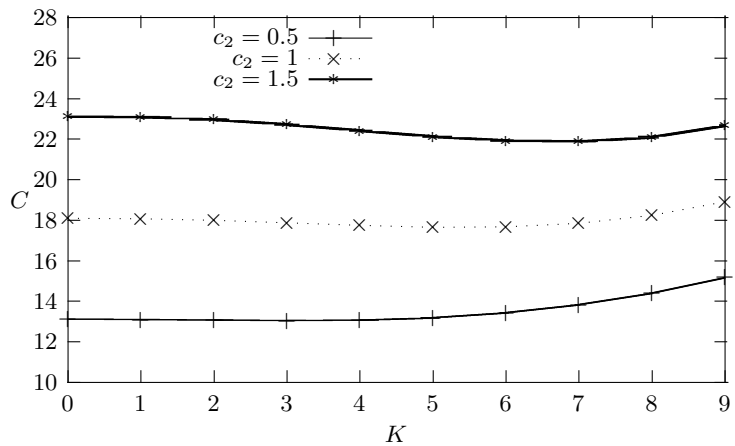

In Figure \ref{f6}, the cost $C$ is plotted against $K$, which is varied 
between 0 and 9. The unit holding cost, is fixed at $c_1=1$, while the
unit energy cost takes three values, $c_2=0.5$, $c_2=1$ and $c_2=1.5$. 
The service time parameters are $\mu_1=1$, $\mu_2=0.2$and $q=0.1$, 
which means that the average service time is 1.5. The job arrival rate is 
$\lambda =5$. Thus, the offered load is 75\% of the maximum system 
capacity.

The figure shows that the threshold policy has a small, but significant 
effect on costs. As one might expect, the larger the energy unit cost, 
the larger the optimal switch-off threshold. Those optima are $K=3$ 
for $c_2=0.5$, $K=5$ for $c_2=0.5$ and $K=7$ for $c_2=1.5$.

\section{Conclusions}

We have introduced and analysed two models of both theoretical and 
practical interest. Exact numerical solutions were obtained in the general 
case, while special cases were solved in closed form. A number of experiments 
were carried out, providing insights that could not have been predicted. In 
particular, it was discovered that the two-phase Foreground-Background 
policy has a better average performance than the optimal LAS policy. Less 
unexpected, but still not obvious, was the observation that for a given 
maximum speed level, it is enough to make an optimal choice of one 
intermediate speed. Optimizing two or more intermediate speeds yields 
only marginal improvements. 

A two-phase approximation for a three-phase system, based on an 
appropriate fusion of the second and third phases, was examined using the 
exact solution of the former and a simulation of the latter. The accuracy of 
the approximation depends on the structure of the three phases and on the 
offered load. 

One could apply exactly the same analysis to a speed modulation policy wth 
different thresholds. For example, speed $s_1$ could operate when the 
number of jobs present is between 1 and $k_1$; speed $s_2$ when that 
number is between $k_1+1$ and $k_2$, etc. There would then be a larger 
set of unknown probabilities, but there would still be enough equations to 
determine them all.

The model with a number of parallel servers enables us to study various 
capacity modulation policies. One such policy, based on a 
switch-off threshold, was analysed in detail and its effect on the cost 
function was examined numerically.

\section*{Appendix}

{\bf Proof of Lemma.}

We start by constructing a sequence of polynomials based on
the principal diagonal minors of the determinant $D(z)$. Define
\begin{eqnarray}
	Q_0(z) & \equiv & 1 \nonumber \\
	Q_1(z) & = & a_0(z) \nonumber \\
	Q_i(z) & = & a_{i-1}(z)Q_{i-1}(z)-\alpha_{i-1}(z)\lambda z Q_{i-2}(z)
	\;;\;i=2,3,\ldots,m-1\nonumber \;.
\end{eqnarray}
$Q_i(z)$ is a polynomial of degree $2i$, so it
has $2i$ zeros. The following properties hold:
\begin{enumerate}
	\item $sign[Q_i(0)] = (-1)^i$, $i=1,2,\ldots,m-1$.
	\item $sign[Q_i(1)] = 1$, $i=1,2,\ldots,m-1$.
	\item $sign[Q_i(\infty)] = (-1)^i$, $i=1,2,\ldots,m-1$.
	\item If $Q_{i-1}(z)=0$ for some $z$, then $Q_i(z)$
	and $Q_{i-2}(z)$ have opposite signs at that point,
	$i=2,3,\ldots,m-1$.
\end{enumerate}

Properties 1-3 follow directly from the form of the relevant determinants. 
At $z=0$ only the diagonal elements remain and they are negative. At $z=1$, 
adding all rows of the relevant minor to the last one also leads to products 
of diagonal elements, which are now positive. At $z=\infty$ the dominant 
term is a product of negative elements. Property 4 follows from the 
recurrence relations.

Starting with $Q_1(z)$, we note that according to
properties 1-3, its two zeros, $z_{1,1}$ and $z_{1,2}$,
are real and satisfy $0<z_{1,1}<1<z_{1,2}$. At those
two points, $Q_2(z)$ is negative, according to
property 4. Hence, the 4 zeros of $Q_2(z)$, $z_{2,1}$,
$z_{2,2}$, $z_{2,3}$ and $z_{2,4}$, are real and lie
in the intervals $(0,z_{1,1})$, $(z_{1,1},1)$,
$(1,z_{1,2})$ and $(z_{1,2},\infty)$, respectively.
Moreover, the sign of $Q_3(z)$ at point $z_{2,s}$
($s=1,2,3,4$), is $(-1)^{3+s}$ for $s=1,2$, and
$(-1)^{s-2}$ for $s=3,4$.

Continuing in this manner, we find that for
$i=3,4,\ldots,m-1$, the $2i$ zeros of $Q_i(z)$, $z_{i,1}$,
$z_{i,2}$, $\ldots$, $z_{i,2i}$, are real and distinct;
the first $i$ of them lie in the consecutive intervals
between points 0, $z_{i-1,1}$, $z_{i-1,2}$,
$\ldots$, $z_{i-1,i-1}$, 1; the second $i$ are in the
intervals between points 1, $z_{i-1,i}$, $z_{i-1,i+1}$,
$\ldots$, $z_{i-1,2(i-1)}$, $\infty$. Moreover, the
sign of $Q_{i+1}(z)$ at point $z_{i,s}$ is $(-1)^{i+s+1}$
for $s=1,2,\ldots,i$, and $(-1)^{s-i}$ for
$s=i+1,i+2,\ldots,2(i-1)$.

The determinant $D(z)$ is given by
\[
D(z) = a_{m-1}(z)Q_{m-1}(z)-\alpha_{i-1}(z)\lambda z Q_{i-2}(z)\;.
\]

We have already seen that $D(1)=0$. Direct evaluation and
the above observations show that $sign[D(0)] = (-1)^m$, and
$sign[D(z_{m-1,s})] = (-1)^{m+s}$ for $s=1,2,\ldots,m-1$.
Therefore, $D(z)$ has a zero in each of the $m-1$ intervals
$(0,z_{m-1,1})$, $(z_{m-1,1},z_{m-1,2})$, $\ldots$,
$(z_{m-1,m-2},z_{m-1,m-1})$. Moreover, $D(z)$ is negative
at $z_{m-1,m-1}$.

Whether there is another zero in the interval $(z_{m-1,m-1},1)$
depends on the value of the derivative $D^\prime (1)$. That
quantity can be obtained in closed form by adding all rows
of $D(z)$ to the last one, dividing that row by $z-1$,
setting $z=1$ and expanding the resulting determinant along
the elements of the last row. This yields, 
\begin{equation} \label{qm1}
	D^\prime(1) = \mu_1^{m-1}(m-1)!\mu_2 (m-\rho_1-\rho_2)
	\left [ \sum_{j=0}^{m-1} \frac{\rho_1^j}{j!}
	+ \frac{m\rho_1^m}{(m-\rho_1)m!} \right ] \;.
\end{equation}

If $m-\rho_1-\rho_2 >0$, then $D^\prime (1)>0$. Hence, for a 
sufficiently small $\varepsilon$, $D(1-\varepsilon)<0$. In that case, 
$D(z)$ is negative on the interval $(z_{m-1,m-1},1)$, i.e. there are 
no other zeros less than 1. The Lemma is established and a normalizeable
solution to the balance equations exists.

\end{document}